\pdfoutput=1
\documentclass[journal=aip,pre,twocolumn]{revtex4-1}
\usepackage{graphicx}
\usepackage{dcolumn}
\usepackage{bm}
\usepackage{multirow} 
\usepackage{cancel}
\usepackage[utf8]{inputenc}

\usepackage{blindtext}
\usepackage{epstopdf} 
\usepackage{amsfonts}
\usepackage{amsmath}
\usepackage{setspace} 
\usepackage[usenames]{color}
\usepackage[dvipsnames]{xcolor}
\usepackage{soul}
\usepackage{xcolor}
\usepackage[normalem]{ulem}
\definecolor{SkyBlueColor}{HTML}{87CEEB}

\begin{document}
	
	\title{\textbf{\Large Chimera state in a neuronal network under the action of a magnetic field}}

    \author{Ga\"el R. Simo}
	\email{simorosain@yahoo.fr} 
	\homepage{Corresponding author.}
	\affiliation{Research Unit Condensed Matter, Electronics and Signal Processing, Department of Physics, Faculty of Sciences, University of Dschang, PO Box 67, Dschang, Cameroon}
	\affiliation{Laboratory of Electrotechnics, Automatics and Energy, Department of Maintenance, Higher Technical Teachers, Training College (HTTC) of Ebolowa, University of Ebolowa,Cameroon}
	\affiliation{MoCLiS Research Group, Dschang, Cameroon}

	\author{Carmel T. Lambu}
	\affiliation{Research Unit Condensed Matter, Electronics and Signal Processing, Department of Physics, Faculty of Sciences, University of Dschang, PO Box 67, Dschang, Cameroon}
	\affiliation{MoCLiS Research Group, Dschang, Cameroon}

	\author{Adamou Dang Koko}
	\affiliation{Laboratory of Electrotechnics, Automatics and Energy, Department of Maintenance, Higher Technical Teachers, Training College (HTTC) of Ebolowa, University of Ebolowa, Cameroon}

	\author{ 	Patrick Herve Fotso Louodop}
	\affiliation{Research Unit Condensed Matter, Electronics and Signal Processing, Department of Physics, Faculty of Sciences, University of Dschang, PO Box 67, Dschang, Cameroon}
	\affiliation{MoCLiS Research Group, Dschang, Cameroon\hspace{2pt}}

	\author{Robert Tchitnga}
    \affiliation{Research Unit Condensed Matter, Electronics and Signal Processing, Department of Physics, Faculty of Sciences, University of Dschang, PO Box 67, Dschang, Cameroon}
	\affiliation{Research Unit CREST - Centre for Research and Engineering in Space Technologies, École Polytechnique de Bruxelles, Université Libre de Bruxelles (ULB)}
    \affiliation{MoCLiS Research Group, Dschang, Cameroon}

	\author{Hilda A. Cerdeira}
	\affiliation{S\~ao Paulo State University (UNESP), Instituto de F\'{i}sica Te\'{o}rica, Rua Dr.Bento Teobaldo Ferraz 271, Bloco II, Barra Funda, 01140-070 S\~ao Paulo, Brazil}
	\affiliation{Epistemic, Gomez $\&$ Gomez Ltda. ME, Rua Paulo Franco 520, Vila Leopoldina, 05305-031 São Paulo, Brazil}

	\date{\today}
	
	\begin{abstract}
        The Hindmarsh-Rose (HR) neuronal network has recently been the subject of studies highlighting the influence of the electric field on  its chimera states. In this study, we demonstrate the influence of the magnetic field on three categories of chimera states previously discovered in the same network: the traveling chimera state, the traveling multicluster chimera state, and the traveling multicluster chimera breather. The study is entirely numerical and, in each case, applies the magnetic field in three ways: first, the entire network is subjected to the field; then, half of the network is subjected to it; and finally, two symmetrical but distinct regions are also subjected to the field. Several phenomena emerge, most notably the multi-traveling chimera state and the multi-alternating chimera state. This illustrates the magnetic field's ability to transform areas of incoherence into areas of coherence, thus enriching the synchronization field and shedding more light on the field’s influence on brain cells.

	\end{abstract}
	
	\maketitle
	
	\textbf{Collective behaviors in neuronal networks arise from interactions between network structure and local dynamics. These behaviors include chimera states, characterized by the coexistence of coherent (synchronized) and incoherent (desynchronized) neuronal populations. This phenomenon is particularly significant as it contributes to our understanding of unihemispheric sleep, epileptic seizures, and information processing in the brain. Although chimera states have been studied in various dynamical systems, their control remains largely unexplored. They have recently been investigated in the context of electric field influence. In this study, we examine the influence of a magnetic field on a network of Hindmarsh-Rose neurons coupled locally via chemical synapses. Our results show that we can induce various behaviors in the network and generate a range of chimera states simply by adjusting the frequency and application of the external magnetic field across multiple zones.}
	\let\thefootnote\relax\footnotetext{\textsuperscript{*}Corresponding author. Email: simorosain@yahoo.fr}
	
	\section{Introduction}
%
Electromagnetic waves, created by several sources that surround us, raise questions and concerns about their impact on health. Many sources of exposure to electromagnetic waves come from the immediate environment (radio, cell phone, etc.), industry (welding equipment, ovens, telecommunications, radars, etc.), or medicine (magnetic resonance imaging examinations, etc.). The rapid evolution of wireless technologies is expected to continue in the coming years, accompanied by their fast-paced diffusion. Humans are continuously exposed to electromagnetic fields throughout their lives \cite{van2009, JuHwan, pall2016}. For instance, while most electrical devices cause transient exposure, some devices emit these fields continuously. This is especially true for electricity transmission lines. Biological and health effects of electromagnetic fields are well defined as a consequence of their interaction with matter \cite{hu2021effects, danho2025, melnick2025, siddiqi2023}. Therefore, it is imperative to focus on how these fields affect the central nervous system.

Neurons form the backbone of the cerebral nervous system. This system alone comprises hundreds of millions of neurons interconnected by chemical and electrical couplings \cite{agosta2013brain,Haoteng}. It thus represents a complex network in which numerous phenomena manifest, such as neurodegenerative diseases \cite{Deramecourt,Seeley,Farina,Juan2014,Christopher,Juan2010}. These phenomena can be analyzed by examining collective behaviors within neuronal populations. One method for analyzing and understanding these collective behaviors is mimicry. To address the problems of mimicry, we use mathematical models of neuronal cells and various network configurations that represent the phenomena under study.

Among several neuronal models, the three-equation Hindmarsh-Rose (HR) model emerged in 1984 \cite{Hindmarsh1984}, offering the possibility of reproducing the three basic activities of a neuron: the resting state, the spike state, and the burst state \cite{Shilnikov, Corson}. This model has evolved over the years, culminating in a four-equation model for greater accuracy \cite{Pinto}, and has also been extended to account for the cell's interaction with the surrounding electric field \cite{Pinto} and/or the surrounding magnetic field \cite{Lv}. The HR model has been used to model one- or two-dimensional neuronal networks to study the collective behaviors that arise within them \cite{jalili2009, corson2009,Qianqian}. Thus, synchronization and chimera states have emerged in the different configurations studied.

Chimera states have been extensively studied in recent decades and continue to generate much discussion. They have been observed in Hindmarsh-Rose neuronal networks in several configurations \cite{Hizanidis, Bidesh, Ram, Gael, Johanne, Fatemeh, Heng, Branislav}. Chimera states are increasingly proving their importance, particularly in neuroscience. Researchers study them not only to understand the behavior of certain aquatic mammals and birds, but also to analyze epileptic seizures \cite{Soumen, Mitchell, Bansal, Zhenhua, Patton}. Today, they are also appearing in other structures such as hypergraphs and swarms \cite{Njougouo, Xinrui,Riccardo, Sarfati, Muolo, lambou}. 
Controlling chimera states remains challenging for those interested in this phenomenon. Nevertheless, some work has addressed this area. This is the case of  Muolo et \textit{al}. \cite{Riccardo}, who discussed pinning control. They applied an external force to the network to modify its dynamics and thus influence the chimera state.

Recently, Simo et \textit{al.} \cite{Simo} established the existence of several chimera states in HR neuronal networks and showed that an external electric field could create several varieties of these states. It therefore becomes pertinent to ask: what contribution an external magnetic field might make? 

The work we address now is a logical continuation of the previous one, in which we contribute to answering the preceding question about the external magnetic field. Specifically, we examine its impact on the three types of chimera states identified: the traveling chimera state, the traveling multicluster chimera state, and the traveling multicluster chimera breather. To do this, we apply the magnetic field in three different ways: first, to the entire network, then to one half of the network, and finally to two symmetrical parts of the network. This is mainly because biological cells in a network may not be exposed to the field all at once.\\
The rest of our work is organized around 2 main points, namely:  Section II talks about the effects of the external magnetic field, in which we present at point (A) the network under study; then comes point (B) with the application of the magnetic field to the whole network, followed by points (C) and (D) which talk respectively about the partial application and the multiple application, and finally a conclusion is presented in Section III.

	\section{Effect of magnetic field}
	\subsection{Network Description}
	As in the previous case, highlighting the influence of the electric field \cite{Simo}, the network used is a ring of HR-type neurons. Each node is modeled using equations proposed by Lv and Ma in 2016 \cite{Lv}. It is also locally coupled by diffusive bonds with the two nearest neighbors and/or coupled non-locally by chemical bonds with the left and right neighbors. We return to this configuration in memory of the wide variety of chimera states it allowed to obtain in previous works. The following equations mathematically describe the network:
		
	\begin{equation}\label{eq.HR4.mt}
	\begin{cases}
	\dot{x}_i=y_i-ax_i^3+bx_i^2-z_{i}+I+J_i+C_i + k_1{x_i}W(\varphi_i)\\
	\dot{y}_i=c-dx_i^2-y_i\\
	\dot{z}_i=r\left[s\left(x_i-x_{i0}\right)-z_i\right]\\
	\dot{\varphi}_i=kx_i - k_2{\varphi}_i+\varphi_{ext}.
	\end{cases}
	\end{equation}
	
	The membrane potential of the $i$-th neuron is $x_i$; $y_i$ relates to fast currents across the membrane, $z_i$ represents slow currents, and $\varphi_i$ is the magnetic flux across the neuron's (or cell's) membrane. $\varphi_{ext}$ and $I$ are the external inputs. The term $k_1x_i{W(\varphi_i)}$ defines a feedback current on the membrane potential when the magnetic flux is changed, and $k_1$ is the feedback gain. The dependence of the electric charge on the magnetic flux is defined by the memory-conductance as follows: 
    \begin{equation}
	 W(\varphi) = \dot{q}(\varphi)= \alpha + 3\beta{\varphi^2} .
	\end{equation}	
	
	 Here, $xW(\varphi)$ comes from the current through the equation:
		
	\begin{equation}
	i = \frac{dq(\varphi)}{dt}=VW(\varphi)=\frac{dq(\varphi)}{d\varphi} \frac{d\varphi}{dt}=k_1xW(\varphi) .
	\end{equation}
	
	V denotes the induced electromotive force. 
	The terms $kx$ and $k_2\varphi$ describe the membrane potential changes induced by the magnetic flux and the leakage of magnetic flux, respectively. The other parameters are set as follow: $a=1$; $b=3$; $d=5$; $r=0.01$; $s=5$; $x_{i0}=-1.6$; ${\alpha}=0.1$ and ${\beta}=0.02$.\\
	For our coupling scheme, we use the one described by Mishra \textit{et al.} \cite{Mishra}, where a set of $M$ (here $M=100$ ) neurons coupled by electrical and chemical synapses is placed on a ring. The electrical coupling is given by: 
	
	\begin{equation}
	J_i = k_3 \sum_{j=i-1}^{j=i+1}\left(x_{j} - x_i \right) \ .
	\end{equation}

	Here we consider only nearest-neighbor interactions, and $k_3$ is the electrical coupling strength. 
	The chemical synaptic coupling is written as 
	\begin{equation}
	C_i = \dfrac{k_4}{2p-2}\left(x_s-x_i\right) \Big( \displaystyle\sum_{j=i-p}^{i+p}\Gamma\left(x_j\right) -\displaystyle\sum_{j=i-1}^{i+1}\Gamma\left(x_j\right) \Big),
	\end{equation}
	
	where $k_4$ is the chemical coupling strength, $x_s = 2$ is the reversal potential,
	$p$ is the number of neighbors connected on each side except for its two closest ones. The sigmoidal nonlinear function $\Gamma\left(x_j\right)$ is used to model the chemical 
	synaptic dynamics  \cite{Somers}:
	\begin{equation}
	\label{eq.Gamma.mt}
	\Gamma\left(x_j\right)=\dfrac{1}{1+\exp \left[-\lambda(x_j-\theta_s) \right]}.
	\end{equation}
	
	The parameter $\lambda=10$ determines the slope of the sigmoidal
	function and $ \theta_s=-0.25$ is the synaptic firing threshold. To respect the ring topology, we choose periodic boundary conditions, $x_{i}=x_{M+i}$, and the initial conditions are given by:
	 $x_i=0.001(i-\dfrac{M}{2})+\zeta_{xi}$, $y_i=0.002(i-\dfrac{M}{2})+\zeta_{yi}$, $z_i=0.003(i-\dfrac{M}{2})+\zeta_{zi}$, 
	where $\zeta_{xi}$, $\zeta_{yi}$, $\zeta_{zi}$ are small random fluctuations. $i=1,...M$. For the remainder of the work, we used the fourth-order Runge-Kutta numerical integration method. The integration covers 100,000 points, 40,000 of which are considered transient. The integration time step is 0.01.
    \\
We choose the external magnetic field in the following simple form:\begin{equation}\label{eq.Bex.mt}
{\varphi_{ext}}=B_{ext}\sin(2{\pi}f_{m}t),
\end{equation} 
where $f_{m}$ and $B_{ext}$ (here  $B_{ext}=1$) are respectively the frequency and maximum amplitude of the external magnetic field.

\subsection{Application of a magnetic field on the total network}
	
\begin{figure}
		
		\includegraphics[width=7.5cm]{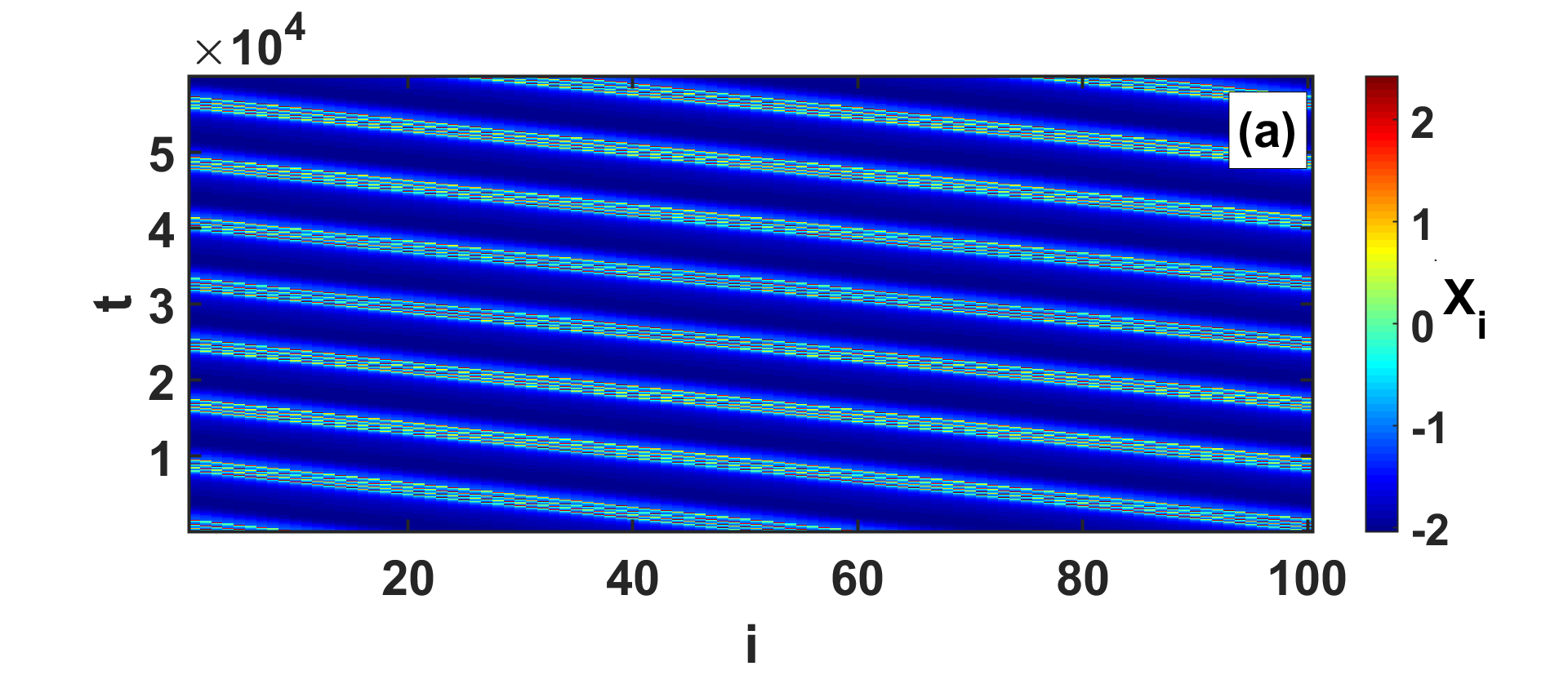}
		
				\includegraphics[width=7.5cm]{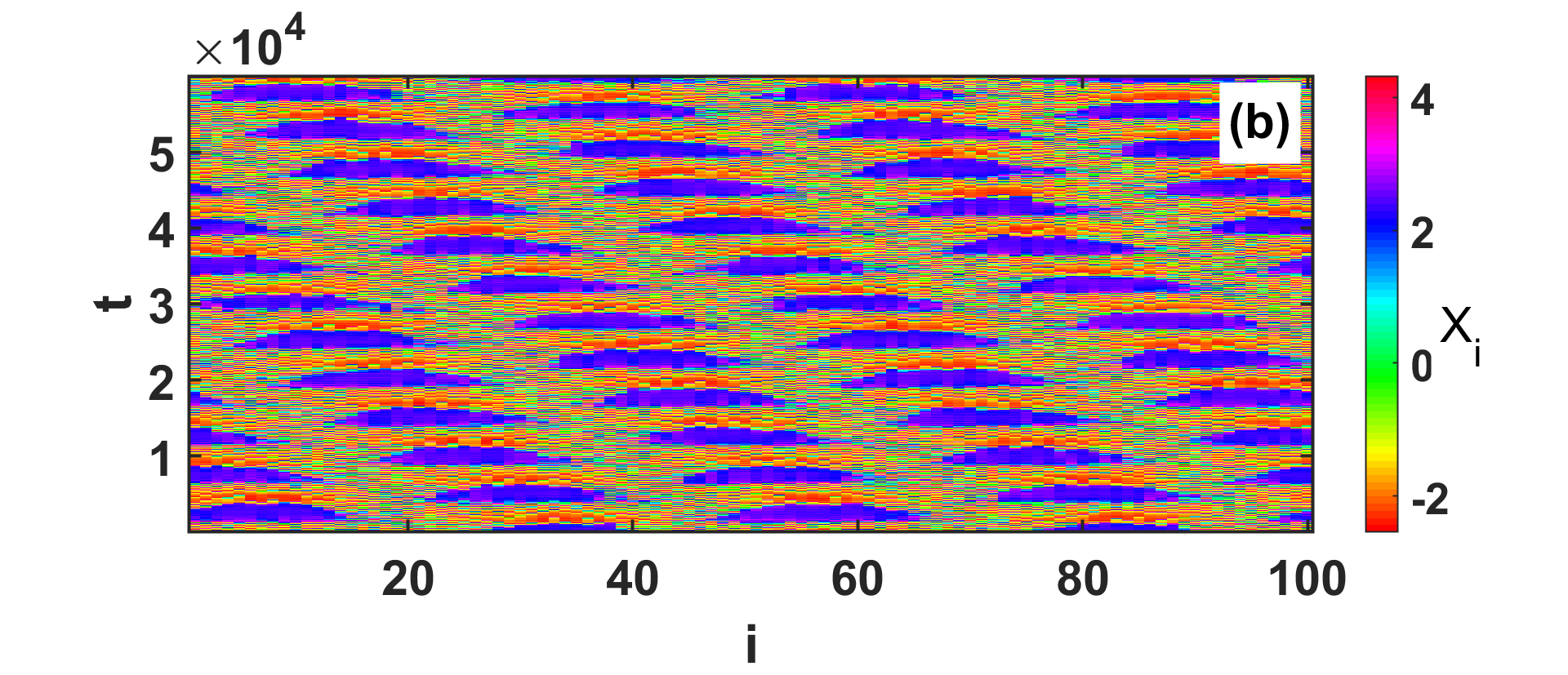}
		
				\includegraphics[width=7.5cm]{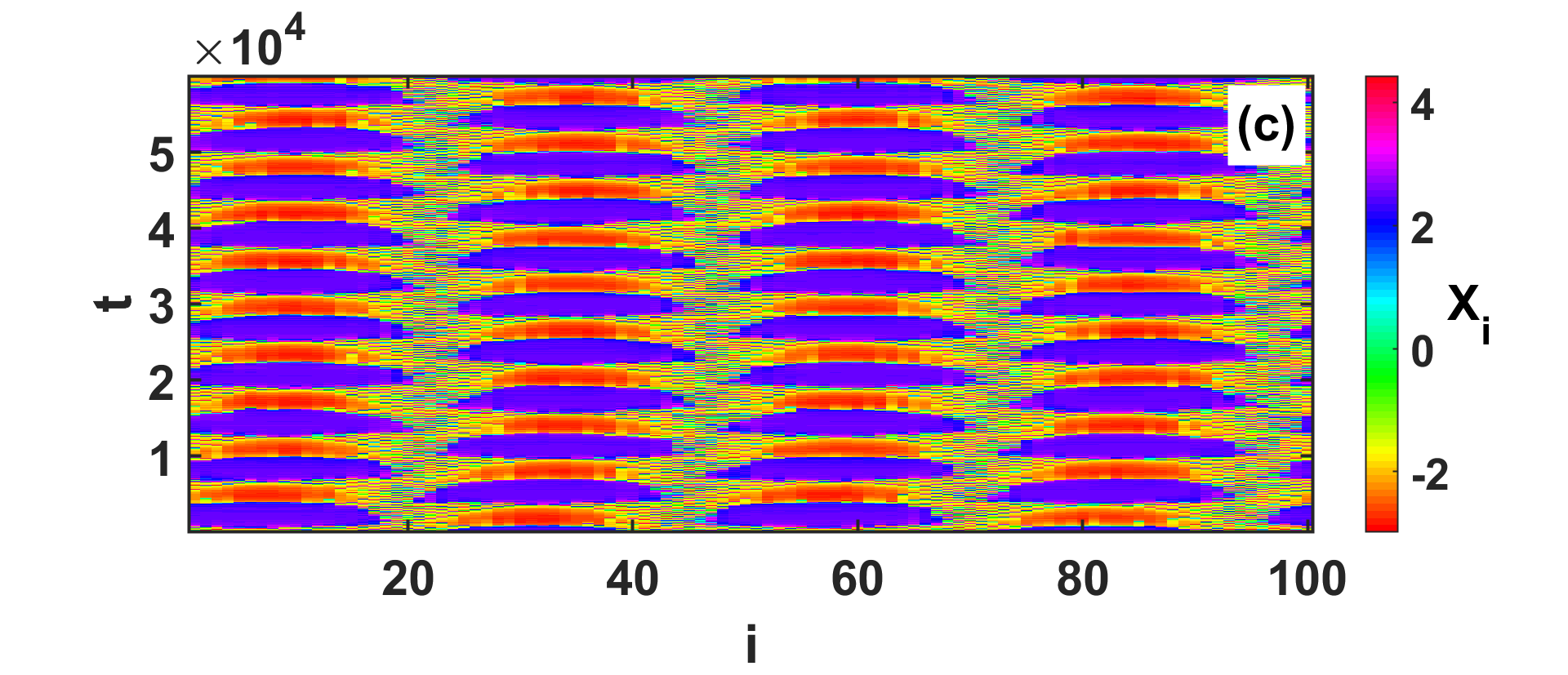}   
            
		\caption{\label{fig.SGR1} 
		 Spatiotemporal evolution of $x_i$ (a)Traveling chimera states, obtained for $k_3=0$, $k_4=9$, $I=3.5$; marked by four coherent regions. (b) traveling multicluster for $k_3=0$, $k_4=9$, $I=35$;  (c)traveling multicluster chimera breathers for $k_3=0$, $k_4=10$, $I=35$. } 
\end{figure}

In accordance with previous work \cite{Simo}, we consider the cases where traveling chimera states, traveling multicluster chimera states, and traveling multicluster chimera breathers were observed; that is, for the following respective pairs of chemical and electrical coupling coefficients: (0,9), (0,10), and for the same excitation current values, namely $I=3.5$ and $I=35$. An illustration for memory recall is provided in Fig.\ref{fig.SGR1}.\\

We begin by applying the magnetic field to the entire network. The process consists of taking the same value for each element. For relatively small frequencies (the case of $f=0.5$), we increase the magnetic coupling $k_1$ from relatively low to high values. 
 \begin{figure}[!h]
		
		\includegraphics[width=7.5cm]{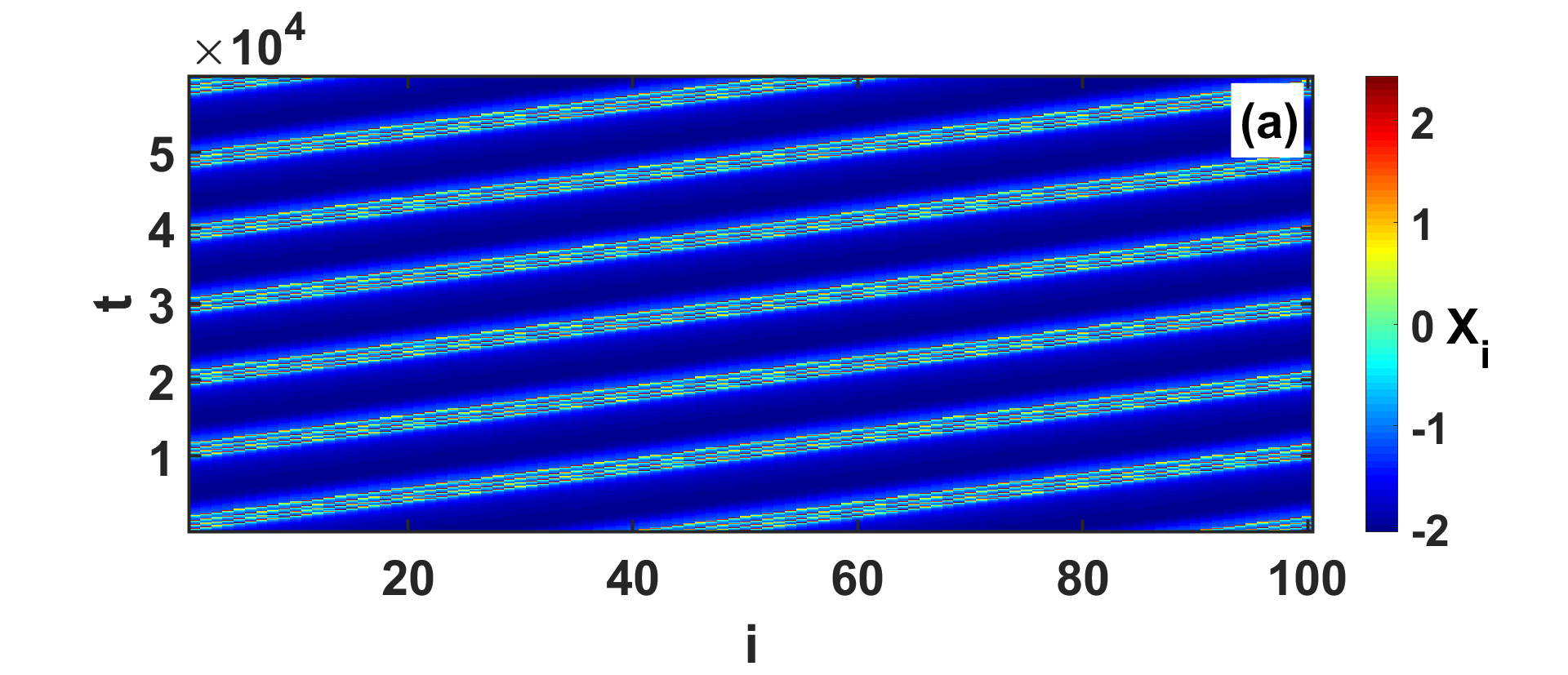}
		
				\includegraphics[width=7.5cm]{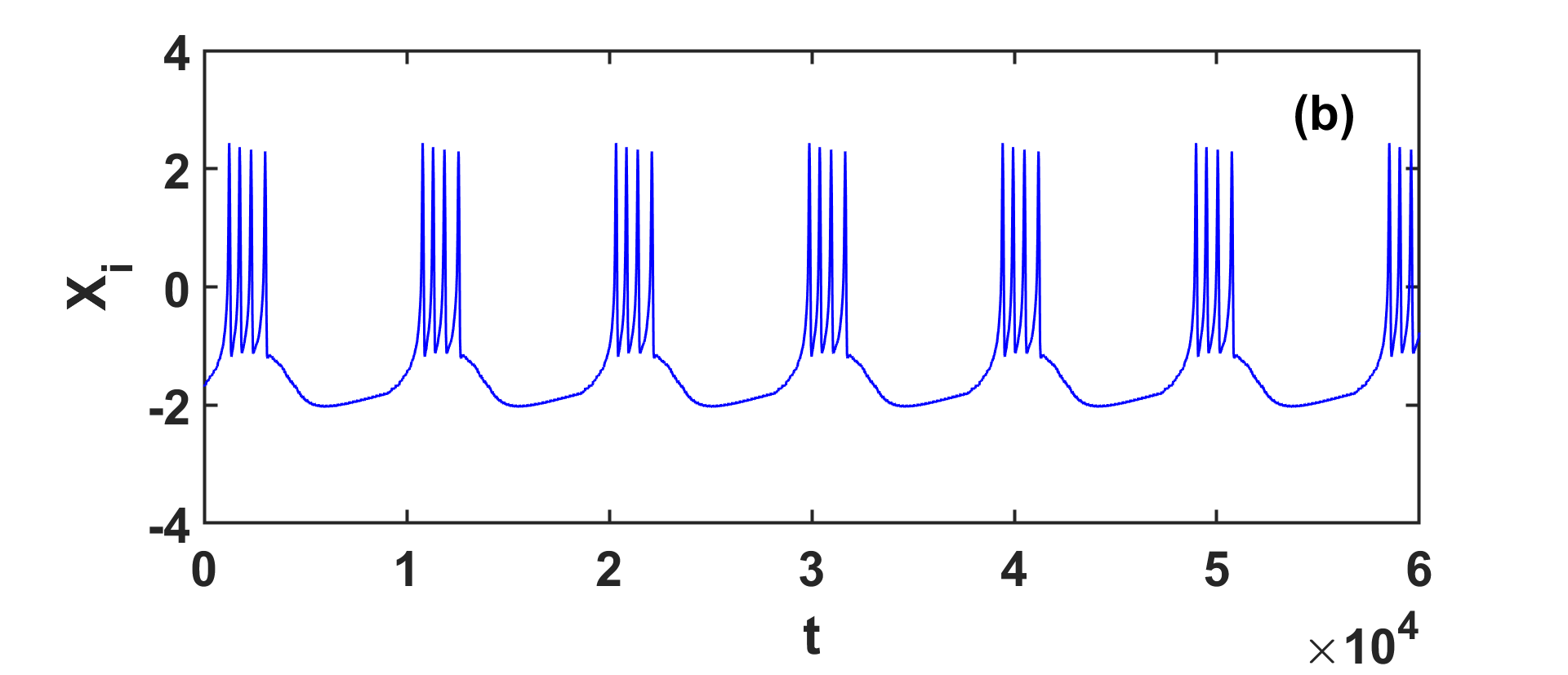}
				
						\includegraphics[width=7.5cm]{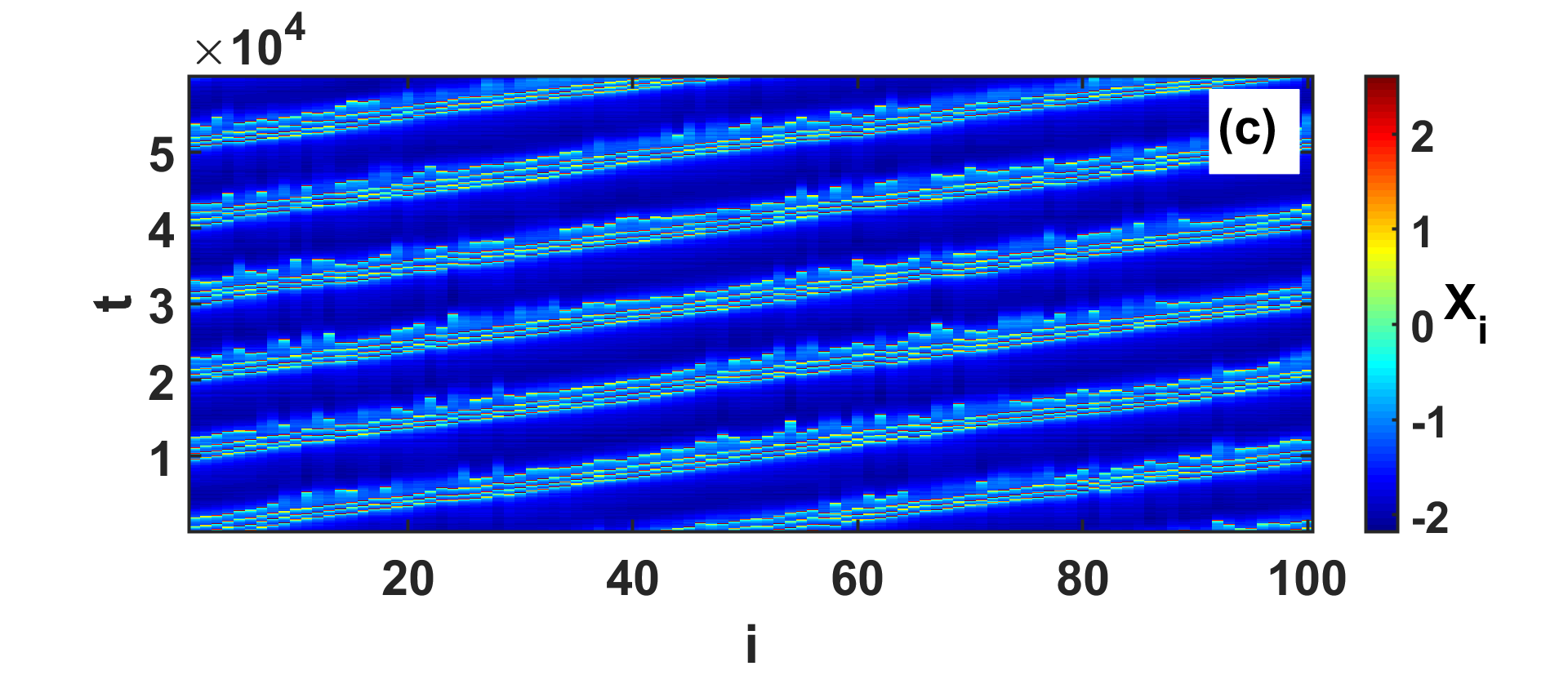}
						
				\includegraphics[width=7.5cm]{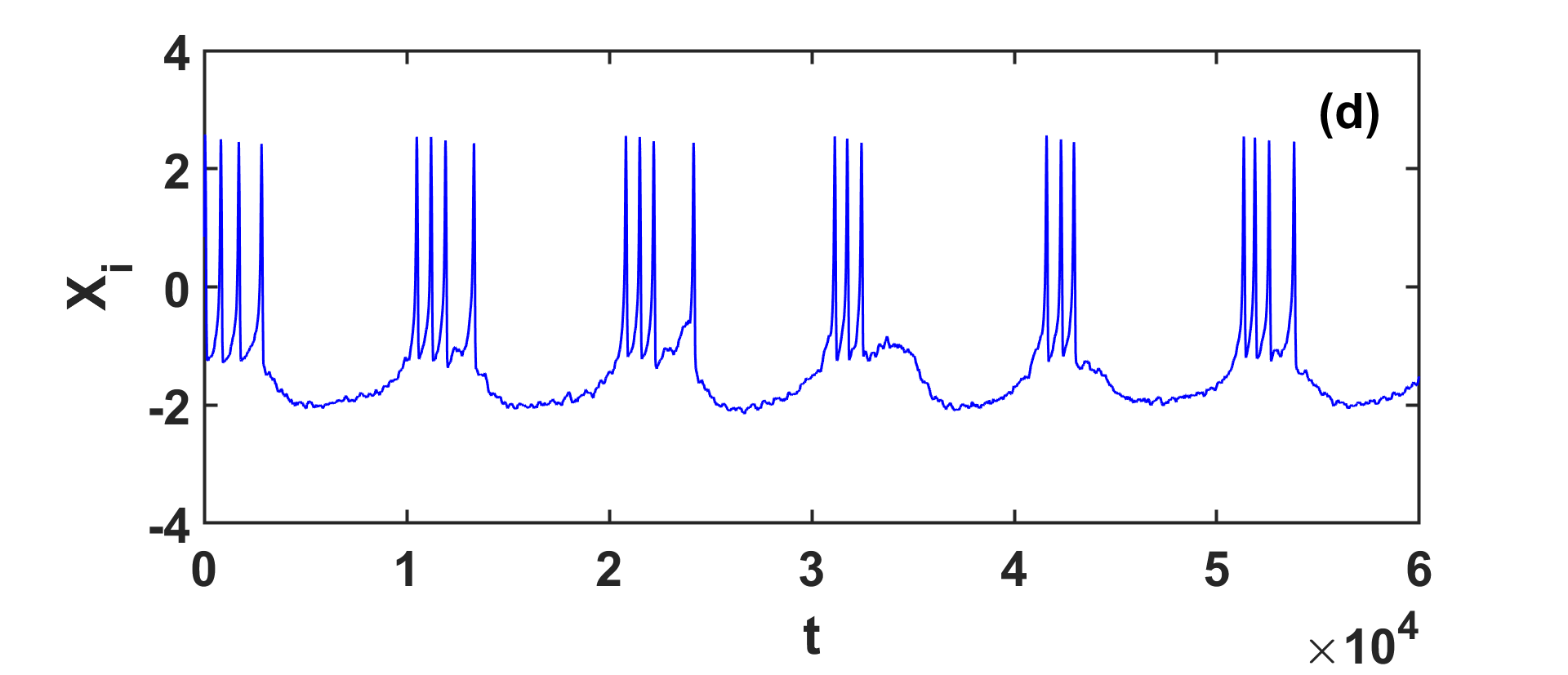}
		\caption{\label{fig.SGR2} 
		 Influence of the magnetic field for $k_3=0$, $k_4=9$ and $I = 3.5$. (a) Spatiotemporal evolution of the $x_i$ for $0\leq k_1<3.5$. A traveling chimera state is observed with (b) regular bursting dynamics on the time series of neurons with every index $i$. 
			(c) Imperfect traveling chimera state for $3.5 \leq k_1<6$ (b)with irregular bursting observed on the time series (d).
			}
	\end{figure}

\begin{figure}[!h]
		\includegraphics[width=7.5cm]{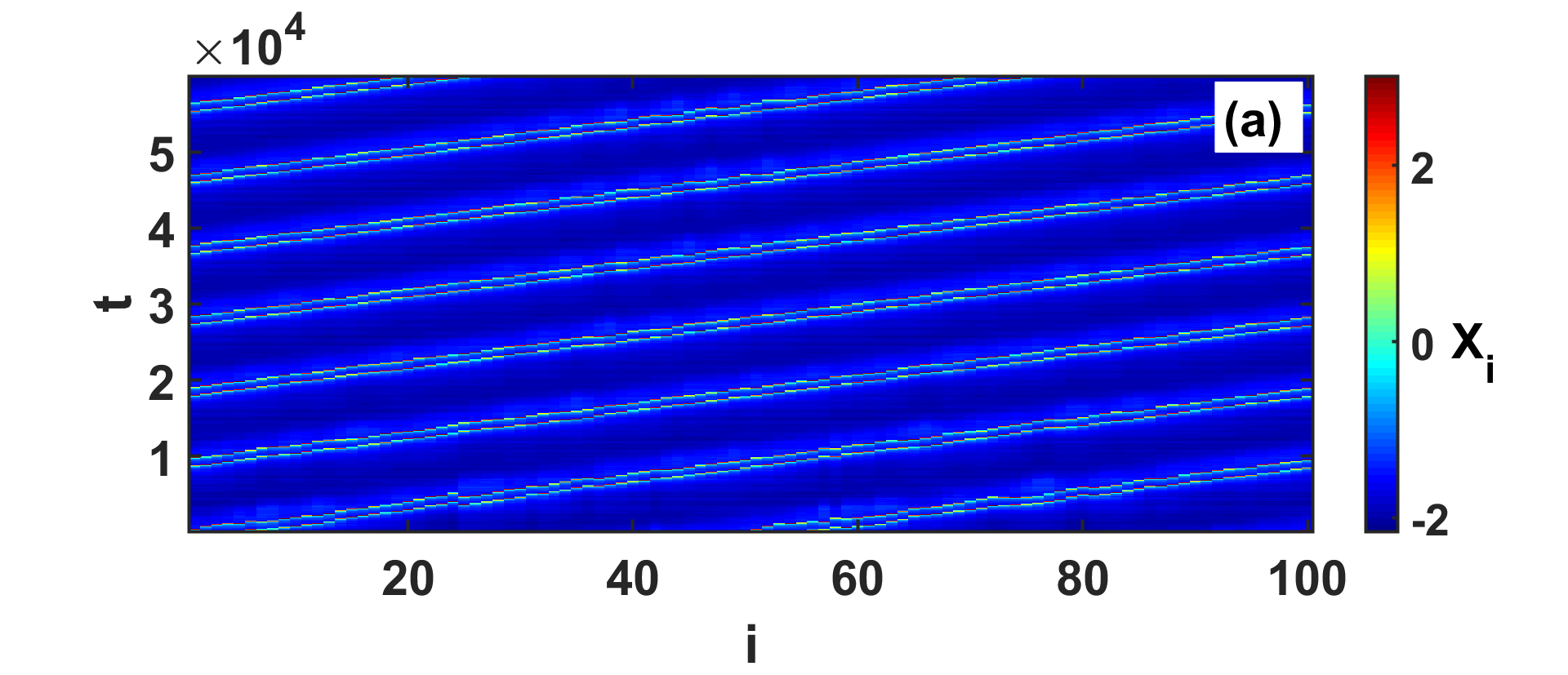}
		
				\includegraphics[width=7.5cm]{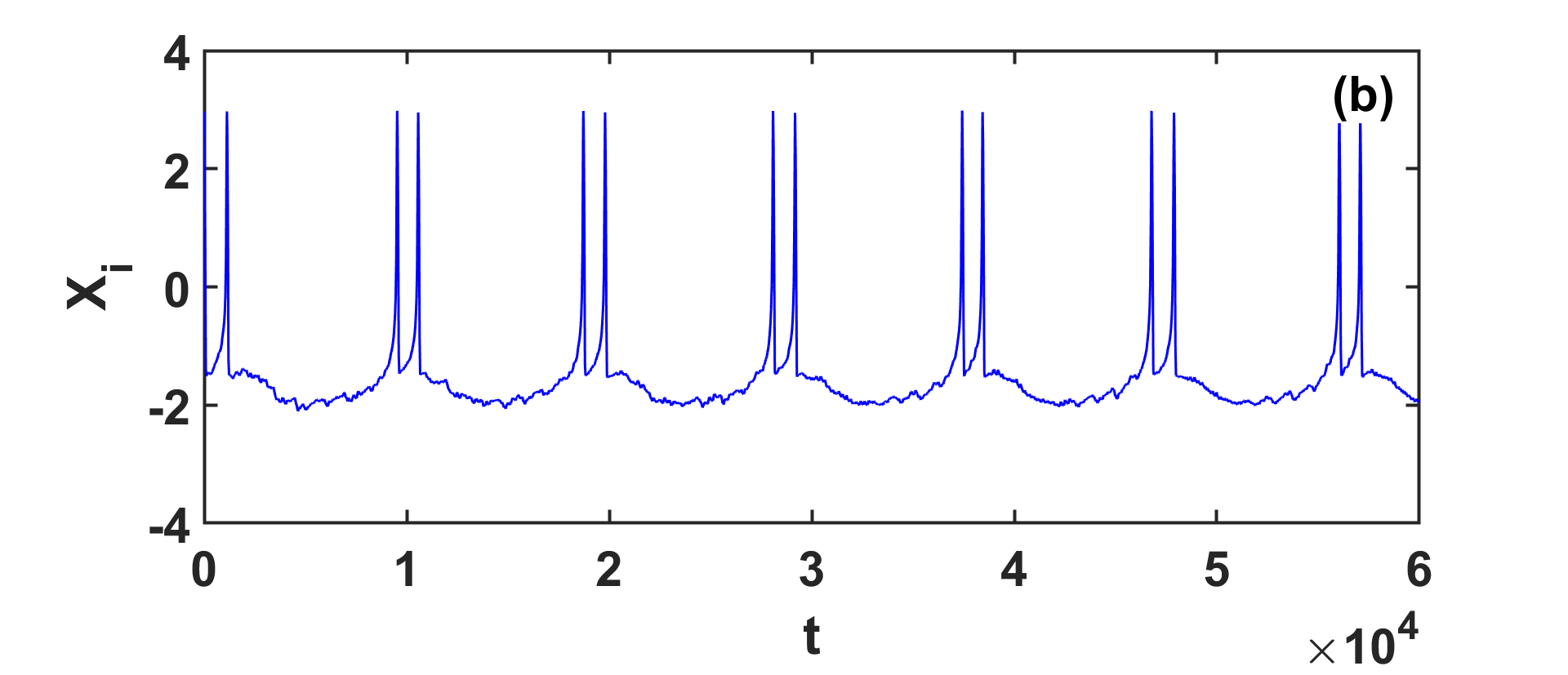}
				
						\includegraphics[width=7.5cm]{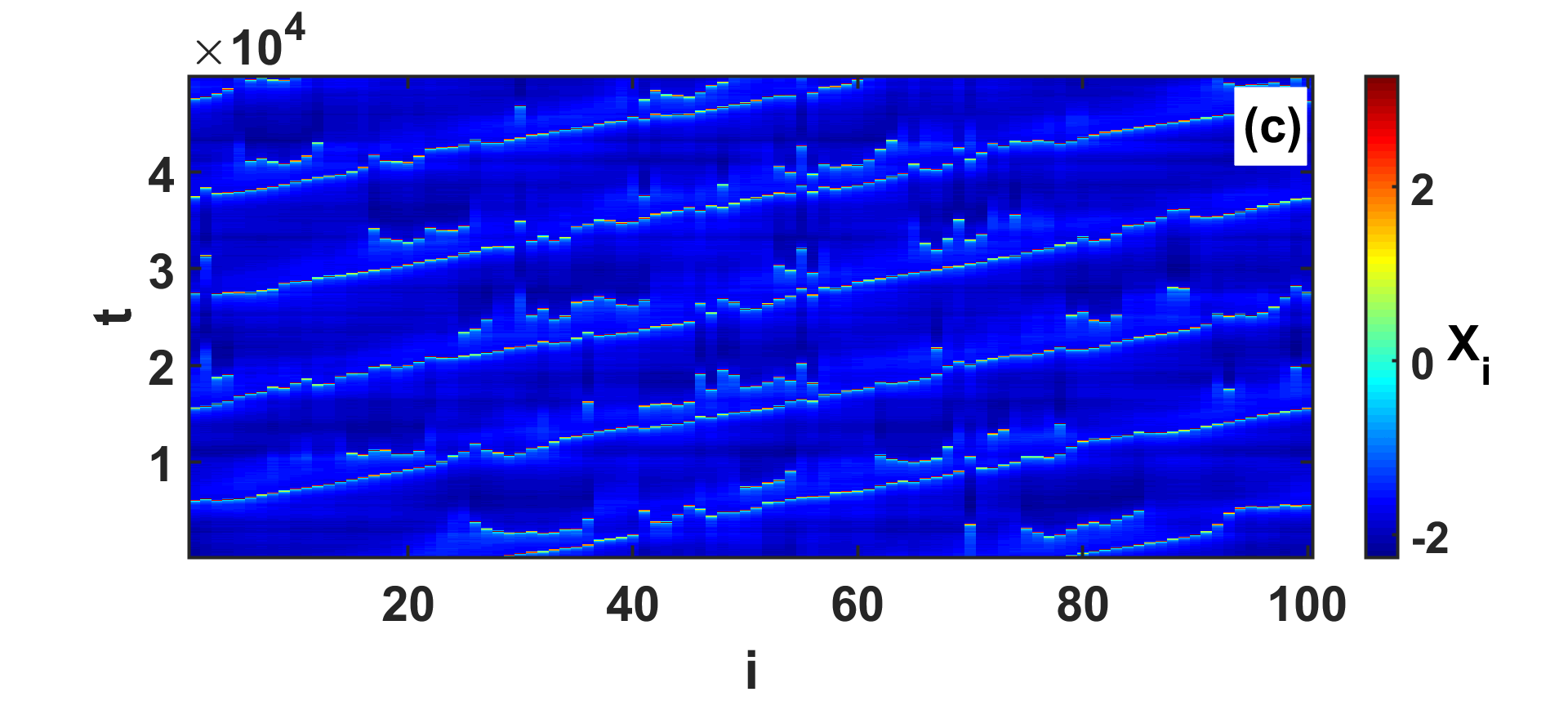}
						
				\includegraphics[width=7.5cm]{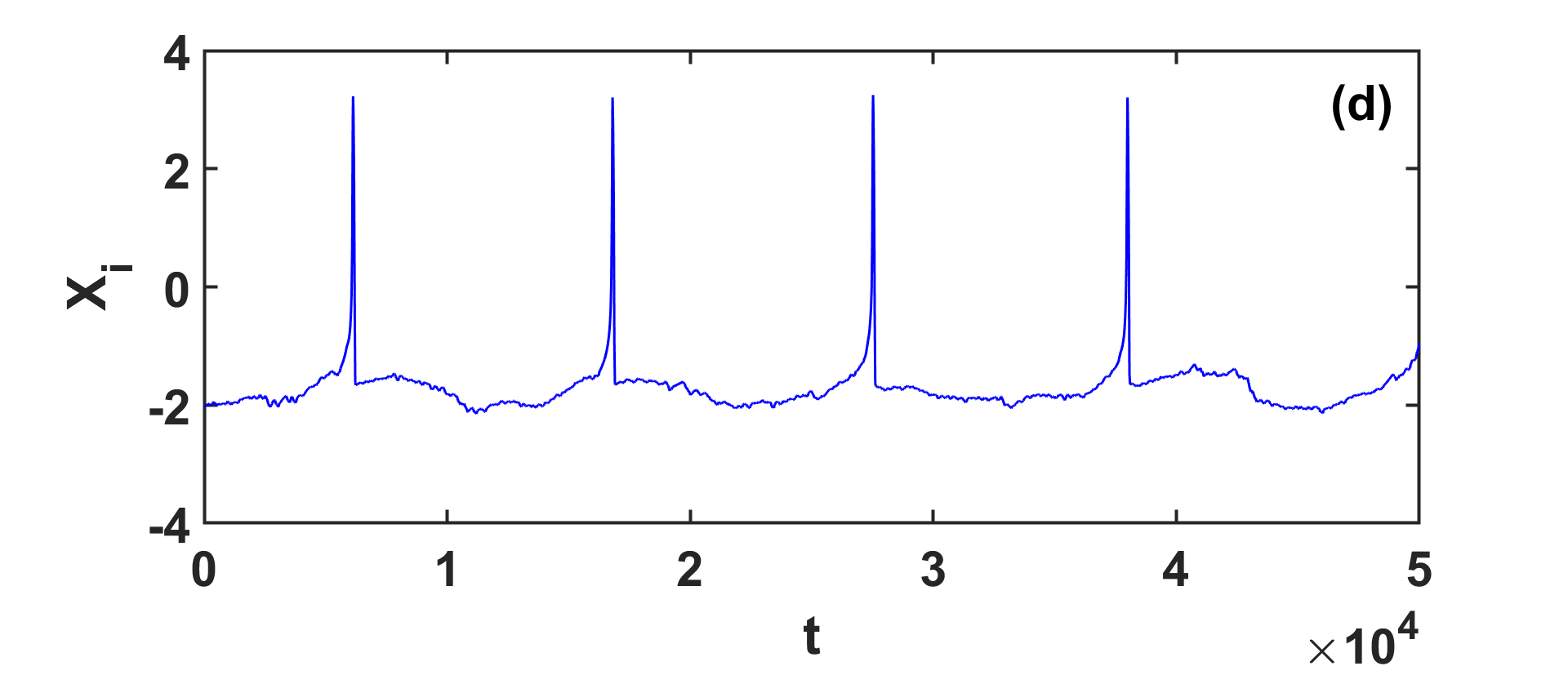}
		\caption{\label{fig.SGR3} 
		 Influence of the magnetic field for $k_3=0$, $k_4=9$ and $I = 3.5$. (a) Spatiotemporal evolution of the $x_i$for $15 \leq k_1<20$. A traveling chimera state is observed in (b), with regular bursting dynamics of two spikes each, in the time series of neurons at every index $i$. 
			(c) Total transformation of the traveling chimera state into an imperfect traveling wave for $k_1 \geq 20$, with spiking observed in the time series in (d).
			}
	\end{figure}

We observe that for the values of $k_1$ between 1 and 3.5 (i.e., $1 \leq k_1<3.5$), the whole network does not change its behavior; it still presents the traveling chimera state as shown in FIG.2(a). Each element of the network has a regular bursting waveform as shown in FIG.\ref{fig.SGR2}(b). For $k_1=3.5$, we observe a degradation of the traveling chimera state. This degradation is materialized by an imperfect traveling chimera state (FIG.\ref{fig.SGR2}(c)) and persists until the boundary of 6 (i.e. $3.5 \leq k_1<6$). Each element of the network manifests irregular bursting as shown in FIG.2(d). When we take the values of $k_1$ between 6 and 11 (i.e., $6 \leq k_1<11$), we see a return to the normal traveling chimera state FIG.\ref{fig.SGR2}(a) and FIG.\ref{fig.SGR2}(b). For the values of $k_1$ between 11 and 15 (i.e., $11 \leq k_1<15$), the imperfect traveling chimera state reappears with irregular bursting as shown in FIG.\ref{fig.SGR2}(c) and FIG.\ref{fig.SGR2}(d). For values of $k_1$ between 15 and 20, we see a return to the traveling chimera state (FIG.\ref{fig.SGR3}(a)), but this time the network elements exhibit regular bursting with two spikes as shown in FIG.\ref{fig.SGR3}(b). Beyond 20, the entire network goes into an imperfect traveling wave state (FIG.\ref{fig.SGR3}(c)) and each element spikes periodically (FIG.\ref{fig.SGR3}(d)). This transformation shows that applying a magnetic field at low frequencies to neurons in this configuration, with strong coupling, can transform a bursting neuron into a spiking neuron.

\begin{figure}
		
		\includegraphics[width=7.5cm]{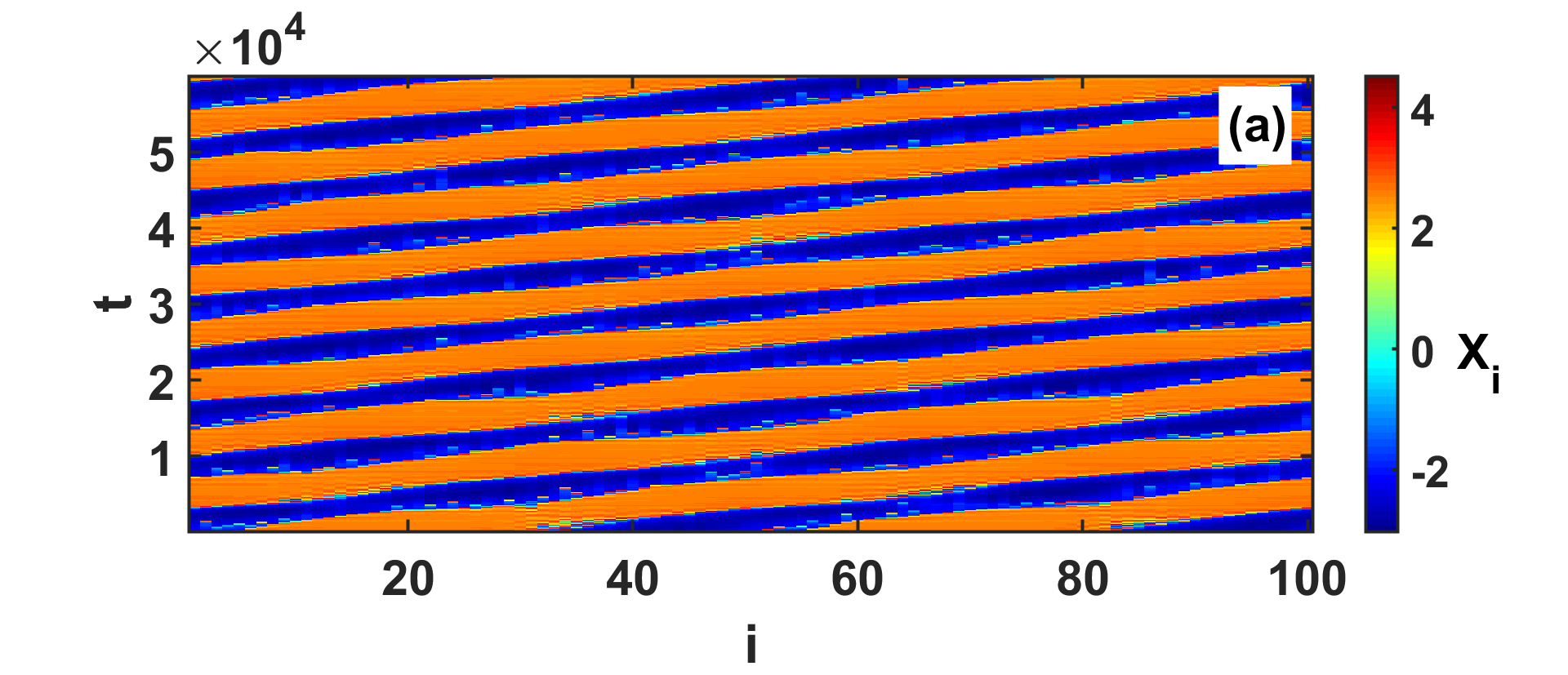}
		
		\includegraphics[width=7.5cm]{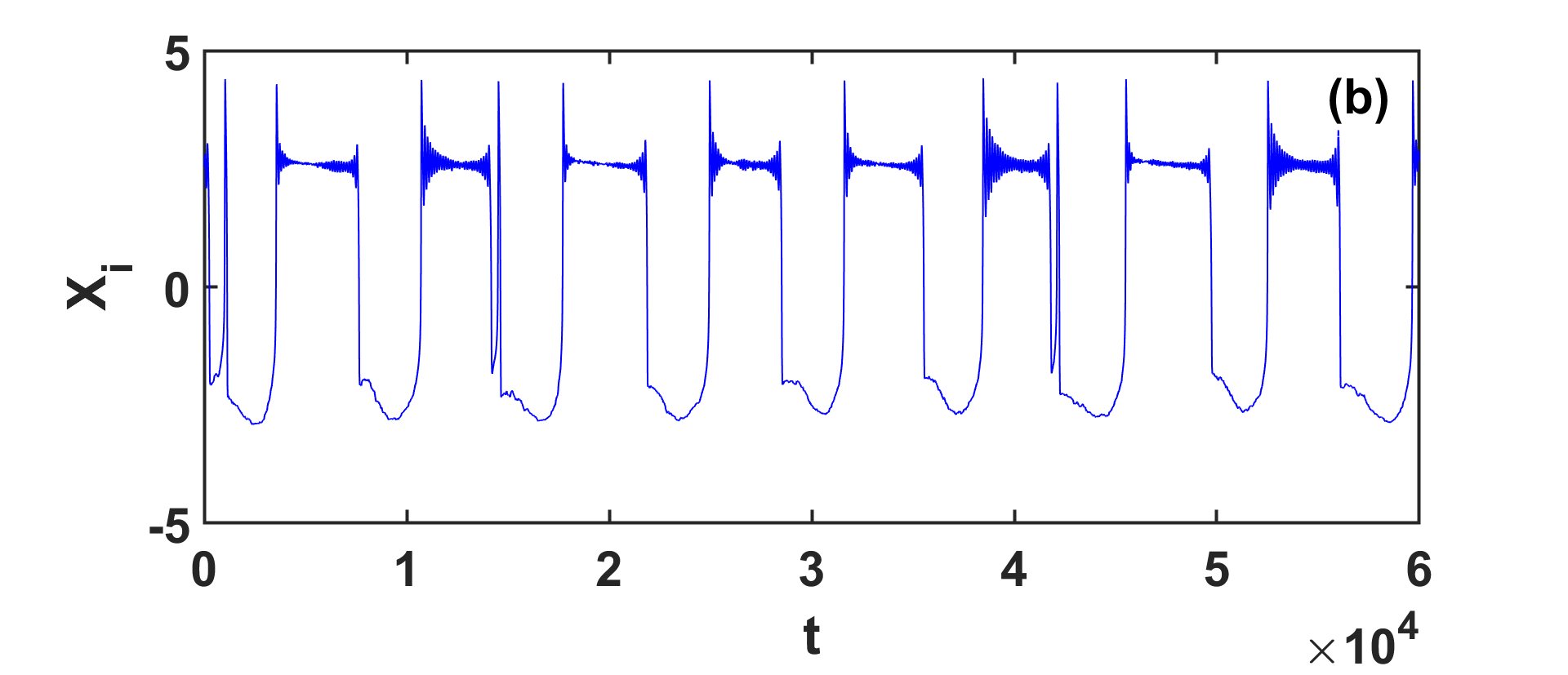}
		\includegraphics[width=7.5cm]{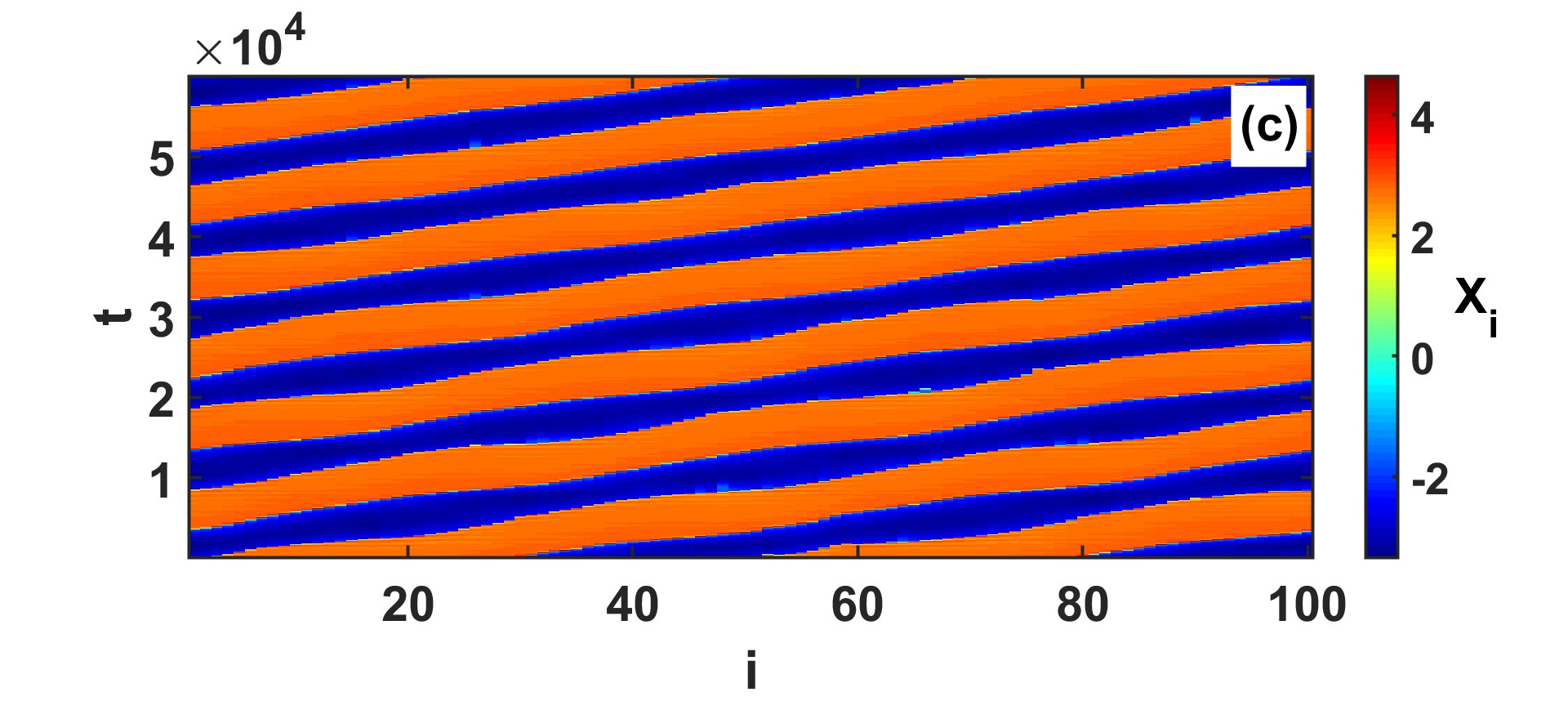}
		
		\includegraphics[width=7.5cm]{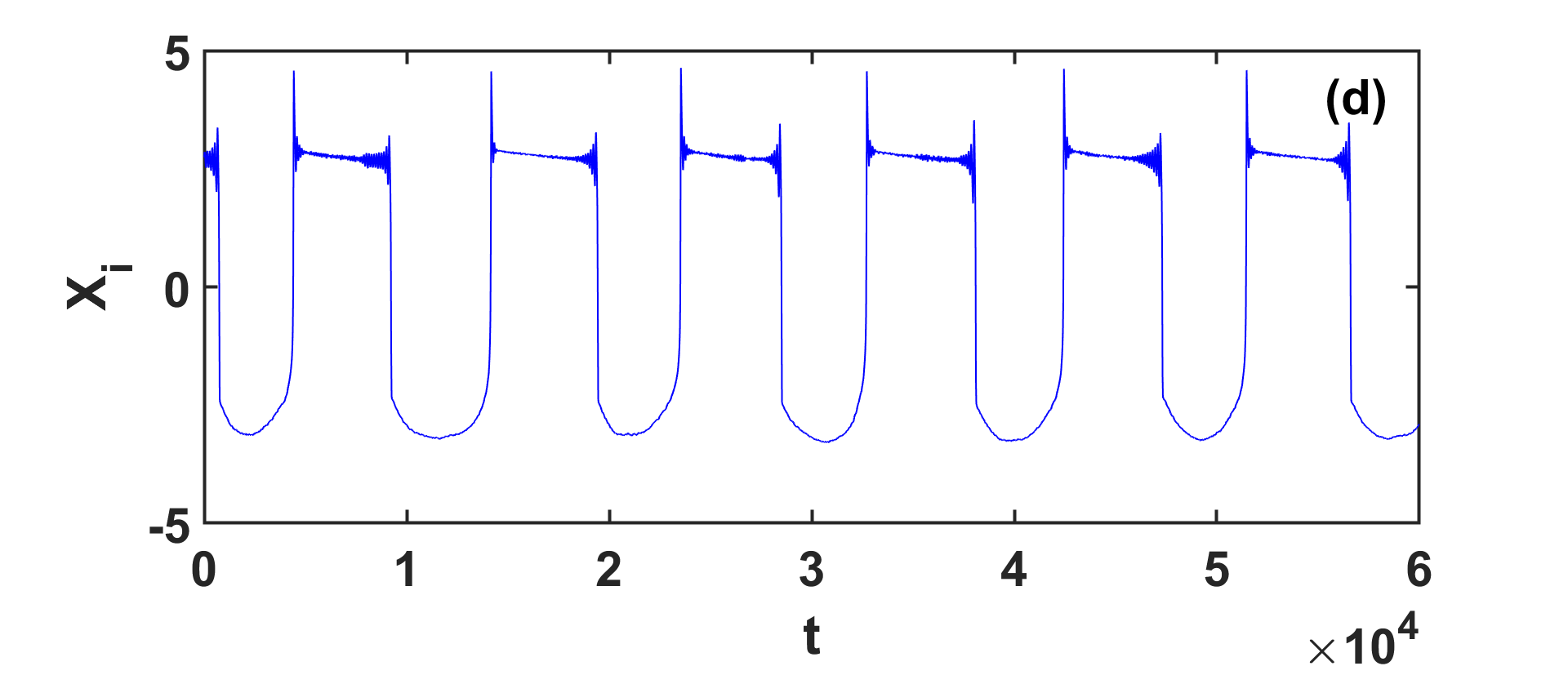}
		
		\caption{\label{fig.SGR4} 
			Influence of magnetic field on traveling multiclusters for $k_3=0$, $k_4=9$ and $I = 35$. (a) Spatiotemporal evolution of the $x_i$for $k_1=8$ showing an imperfect traveling wave, which becomes clear in (b), where the irregular oscillation dynamics as a function of time is shown for a given neuron of index $i$. 
			(c) Traveling wave for $k_1=15$ (b)with regular oscillations observed in the time series.}
		
	\end{figure}
After applying the magnetic field to the traveling chimera state, we now apply it to the traveling multicluster chimera state and to the traveling chimera breathers shown in Fig. \ref{fig.SGR1}(b)  and \ref{fig.SGR1}(c), respectively. For the traveling multicluster chimera state, we observe only two states within the same range of $k_1$ values previously examined. First, the traveling multicluster transitions to the imperfect traveling wave state observed in Fig.\ref{fig.SGR4}(a). This acts as a transient phase, where we have irregular oscillations of each element (Fig.\ref{fig.SGR4}b). Then, the imperfect traveling wave transitions to the traveling wave proper, shown in Fig.\ref{fig.SGR4}(c), where the time series of each individual is a regular oscillation. Regarding the individual behavior in this case, each element of the network exhibits regular oscillations at broad pulse widths, as shown in Fig.\ref{fig.SGR4}(d). Next, we consider traveling multicluster chimera breathers in the presence of a magnetic field, where we observe the same behavior, with the only difference that the transition to the traveling wave starts at 2.5 instead of 8, as in the case of the traveling multiclusters. This indicates that phase transition is much faster for traveling multicluster chimera breathers.

\subsection{Partial Application of  a magnetic field}
\begin{figure}
	
	\includegraphics[width=8.5cm]{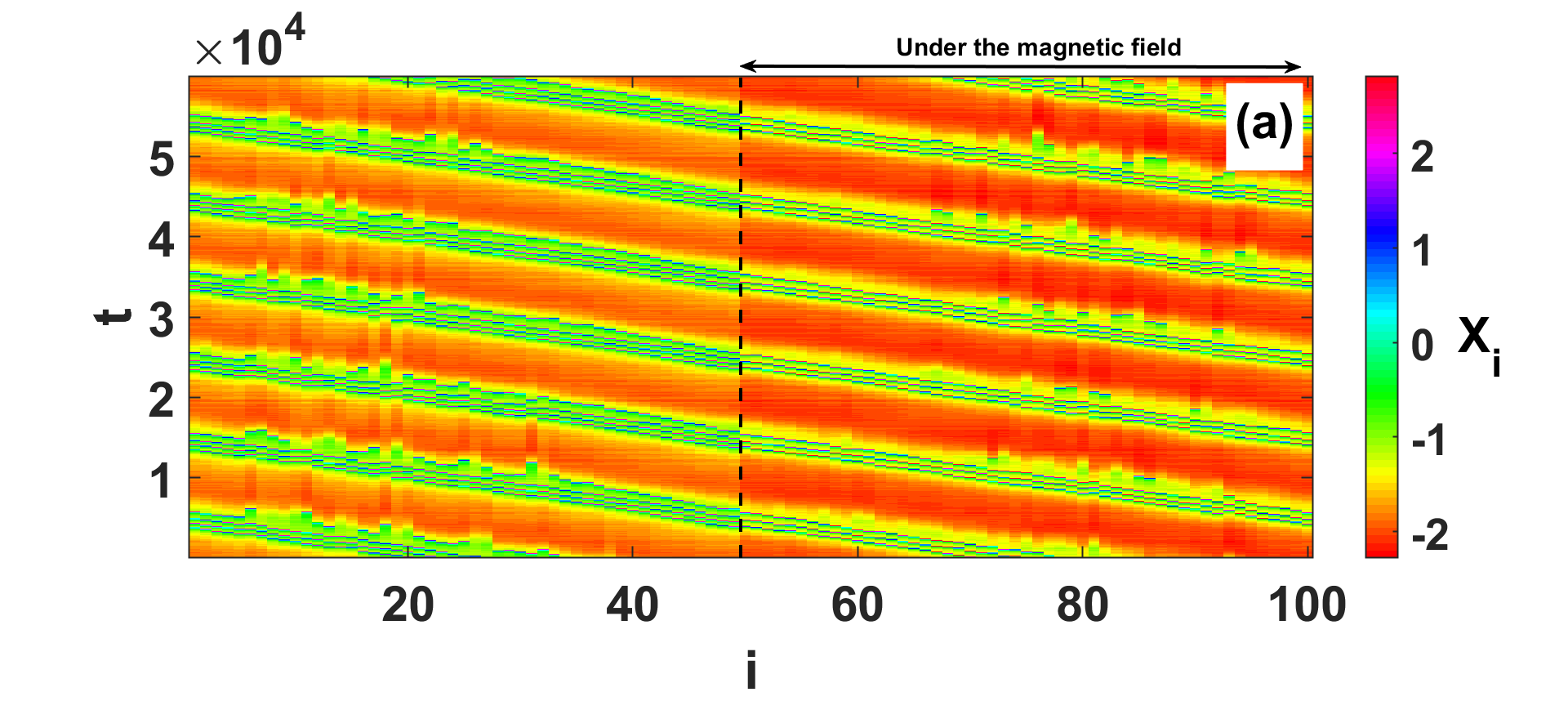}	
	\includegraphics[width=8.5cm]{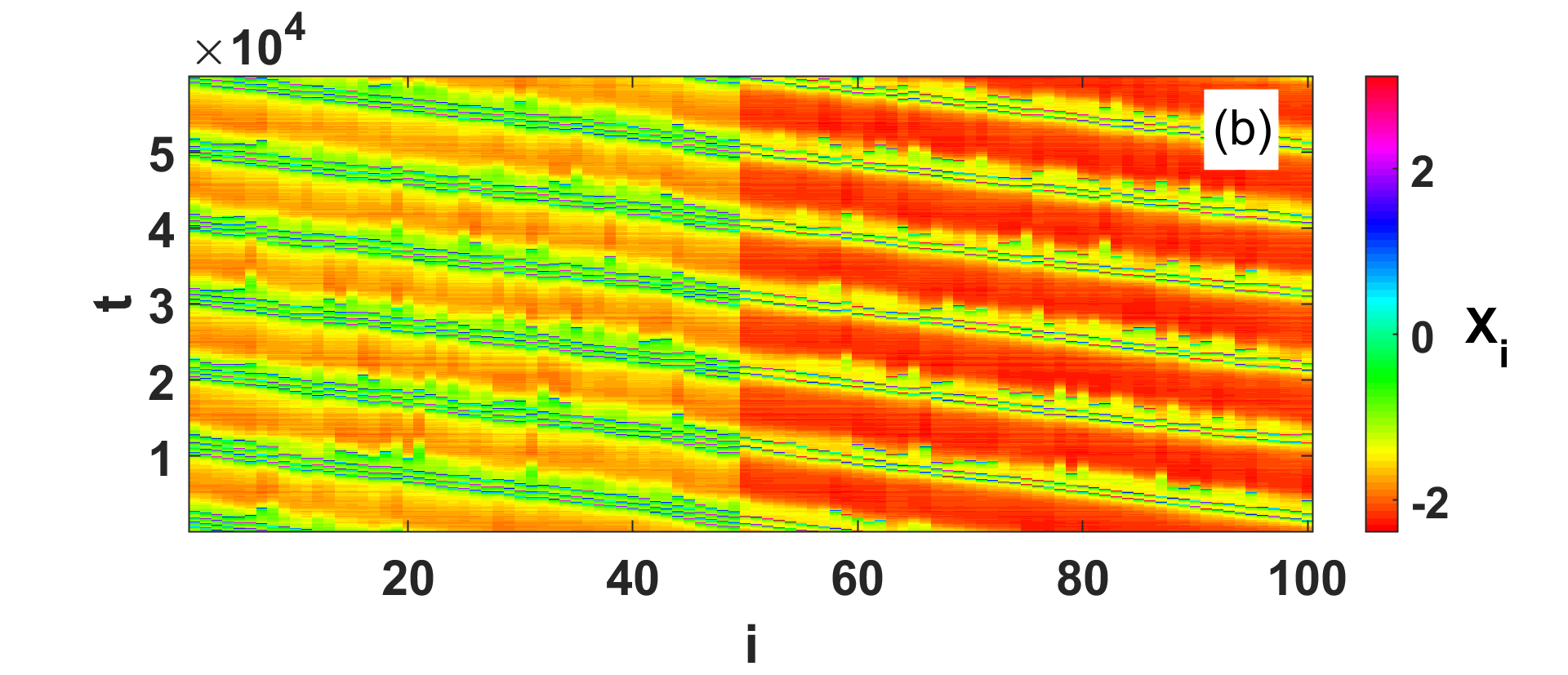}
	
	\includegraphics[width=8.5cm]{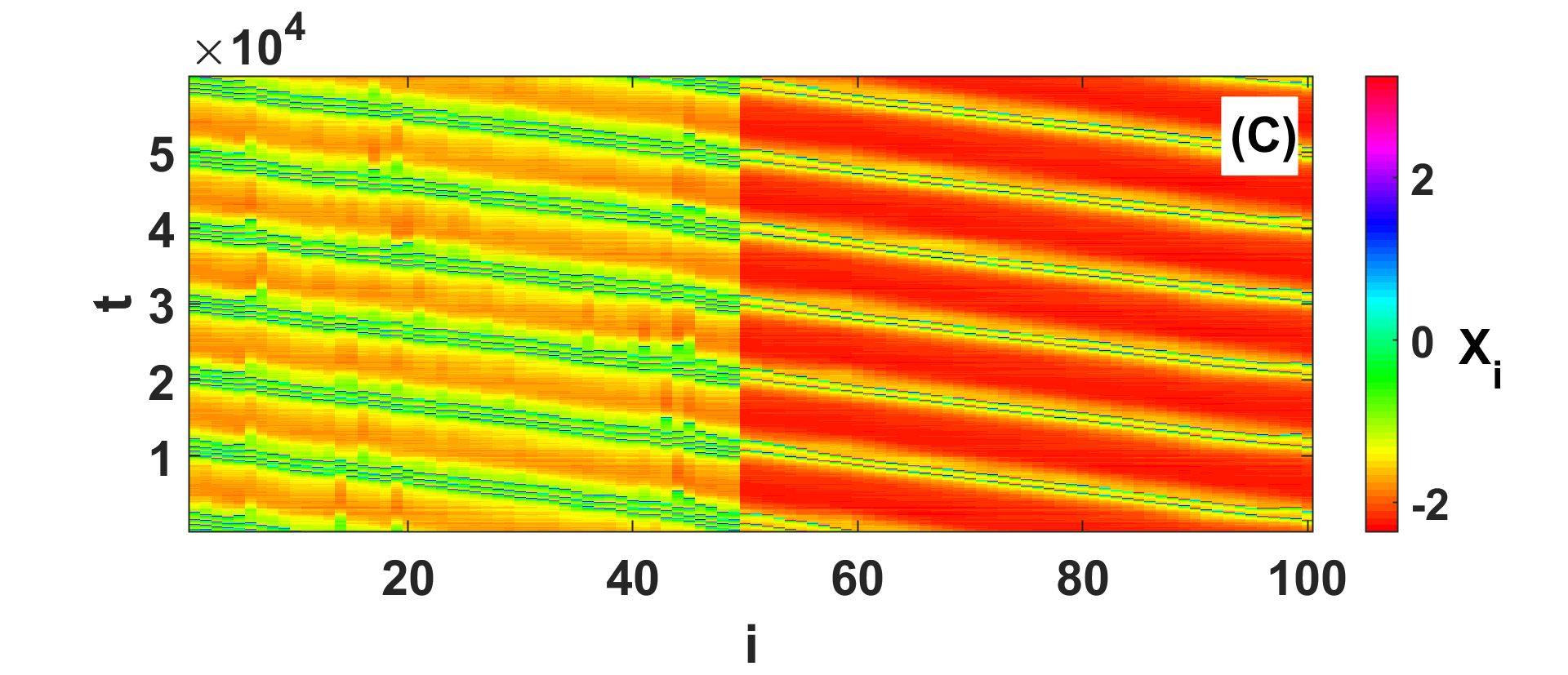}
	\caption{\label{fig.SGR5} 
		Influence of magnetic field on traveling chimera state, applied to half of the network: For $f=0.5$: Spatio-temporal evolutions of the $x$ variables (a) highlighting the degradation of traveling chimera state on both sides of the split line($k_1=6$); (b) appearance of imperfect traveling chimera state on both sides ($k_1=13$); (c) Return to the traveling chimera state in the part subjected to the field ($k_1=20$).}

\end{figure}

We maintain the external field at the current frequency and apply it to a portion of the network, specifically the second half. To do this, we perform this process for values of $k_1$ taken from each of the previously used intervals. We begin with the case of the traveling chimera, where we observe the following: Assigning values less than 8 to the coefficient $k_1$ ($k_1\leq8$) reveals a simple split line in the network, marking a slight vertical shift from one block to the other, as observed in Fig.\ref{fig.SGR5}(a). In most cases, this line is accompanied by the onset of degradation of the traveling chimera state on both sides. When we vary the values of $k_1$ between 8 and 16 ($8 < k_1 \leq 16$), two facts are visible (Fig.\ref{fig.SGR5}b): the appearance of the imperfect traveling chimera state in both blocks and a reduction of the incoherent bands in the area subjected to the field. Beyond 16 ($k_1 > 16$), the field-affected area exhibits a traveling chimera with reduced incoherent bands, automatically increasing coherent bands compared to the regular case. In contrast, the unsubjected area exhibits an imperfect traveling chimera state with incoherent zones maintaining practically the same width. This is shown in Fig.\ref{fig.SGR5}c.

\begin{figure}[!h]	
	\includegraphics[width=7.5cm]{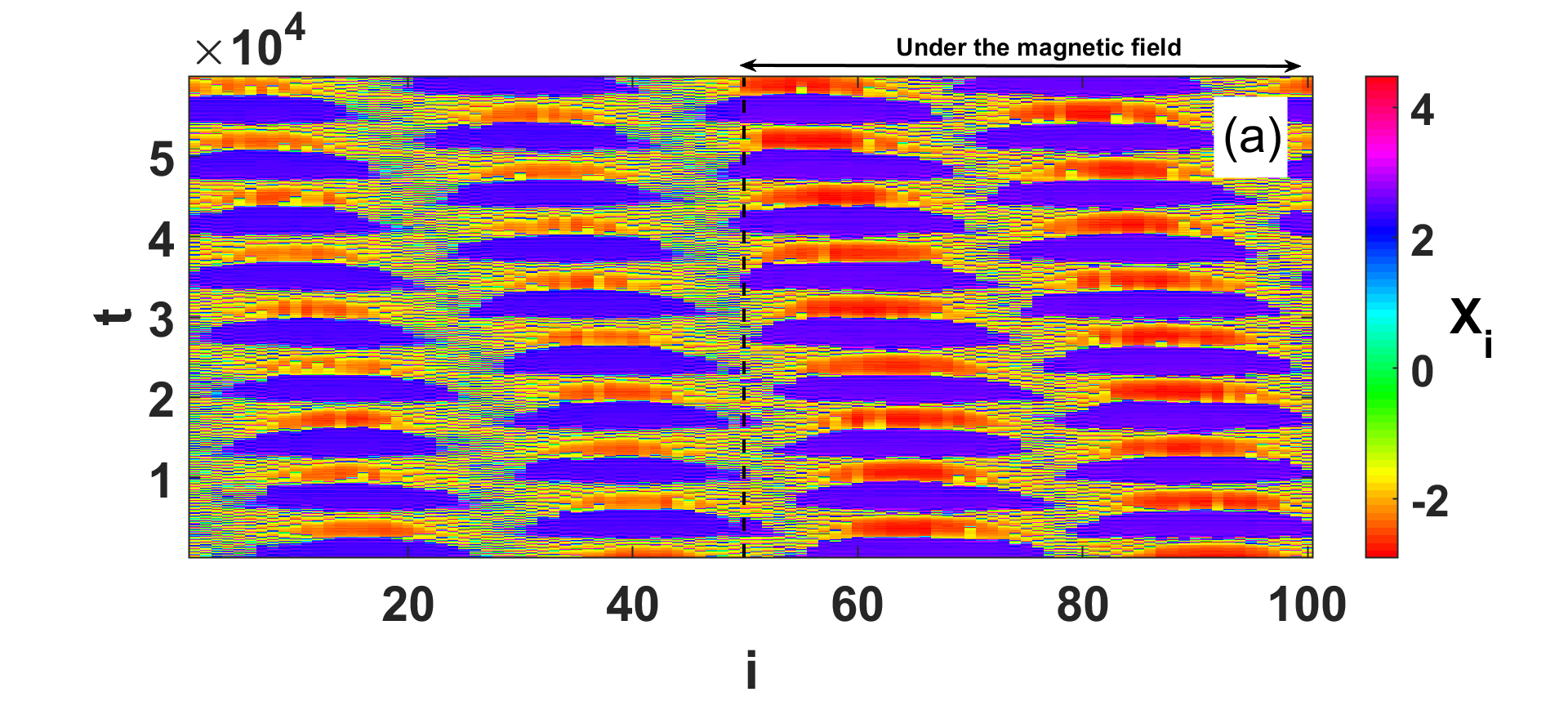}	
	\includegraphics[width=7.5cm]{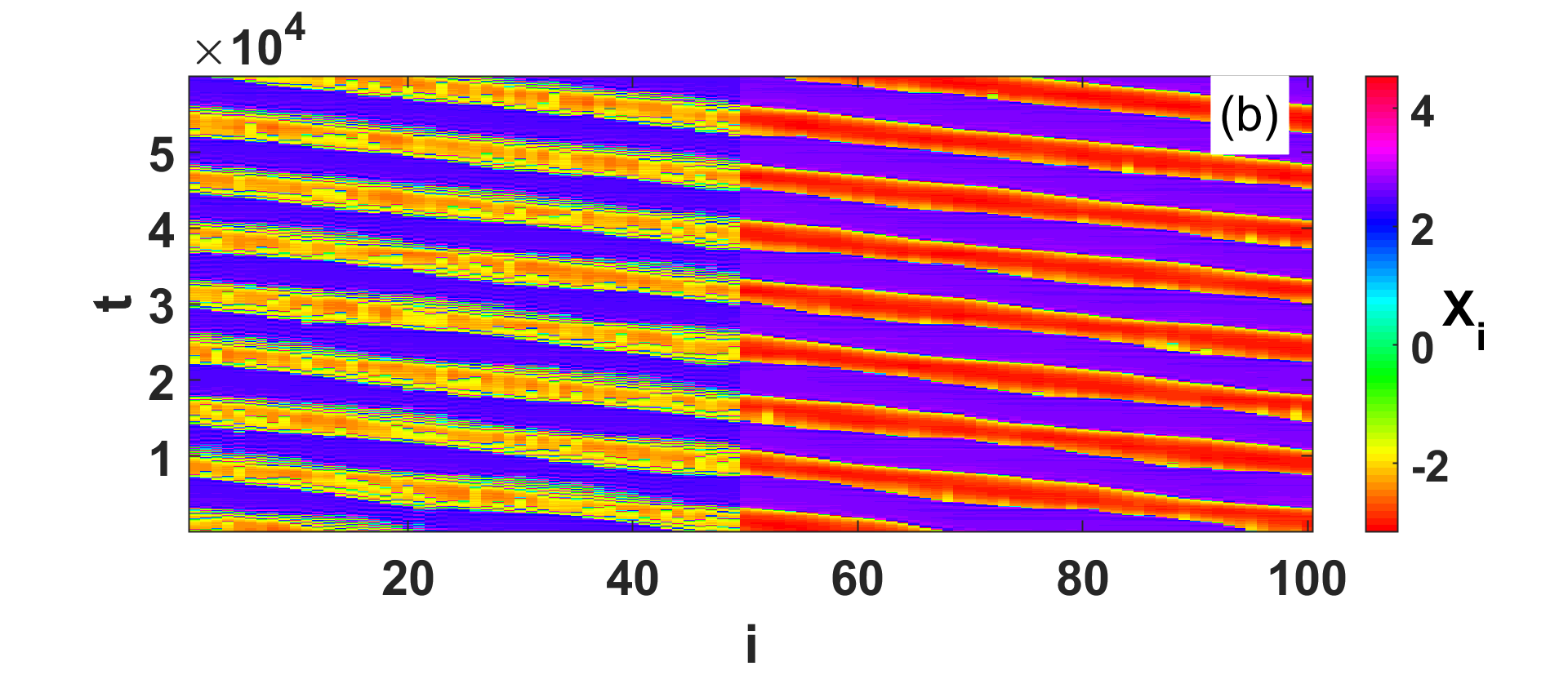}	
	\includegraphics[width=7.5cm]{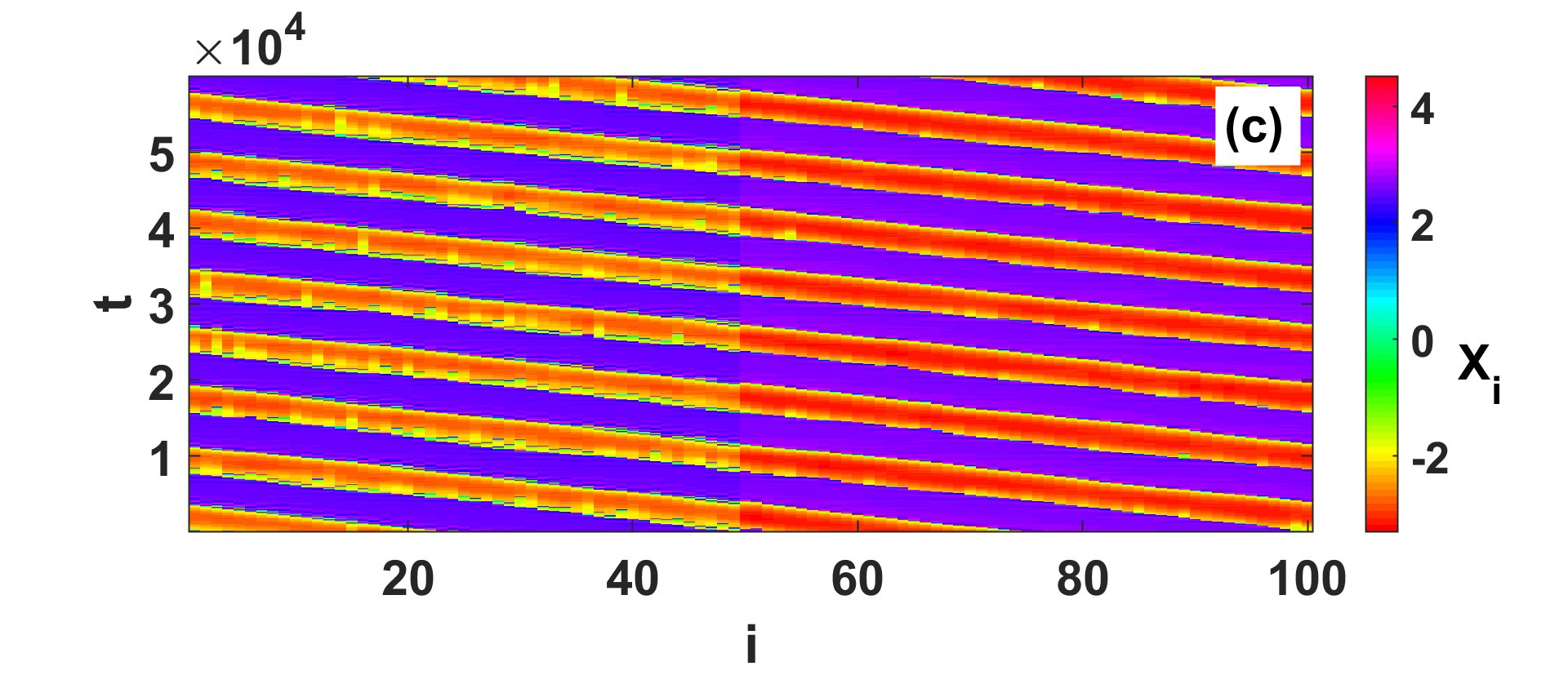}
	\caption{\label{fig.SGR6}
		Influence of the magnetic field applied to half of the lattice for $f=0.5$: Spatiotemporal evolutions of the variables $x$; for traveling multicluster chimera state: (a) for $k_1=8$, a strengthening of the coherence degree on the part subjected to the field; (b) for $k_1=15$, the simultaneous presence of the traveling wave and a traveling chimera state. For the traveling multicluster chimera breather, (c) a traveling wave is formed for $k_1 \geq 4$ ( here is the case where $k_1 = 10$).} 
\end{figure}

Now, we consider the application to the traveling multicluster chimera state. Here, we note that below 8 ($k_1\leq 8$), no major changes are observed in the overall structure. The only difference is that in the area subjected to the field, a strengthening of the colors of the coherent areas is observed, suggesting an increase in the level of coherence (Fig.\ref{fig.SGR6}a). Above 8 ($k_1> 8$), we observe a division of the network into two blocks with different behaviors: the block subjected to the field transitions from multicluster to the traveling wave, marked by the change of incoherent bands into coherent bands (Fig.\ref{fig.SGR6}b), and the block not subjected to such field transitions. This shows that the field can shift the network from a common collective behavior (traveling multicluster) to a duality of behaviors: the traveling wave and the traveling chimera state.

Regarding the traveling chimera breather, it should be noted that when the magnetic field is applied, the entire network goes directly into a traveling wave state for values greater than 4 (Fig.\ref{fig.SGR6}c) and remains in its initial state for values less than or equal to 4.

\subsection{Multiple Applications of  a Magnetic Field}
\begin{figure}[!h]	
	\includegraphics[width=7.5cm]{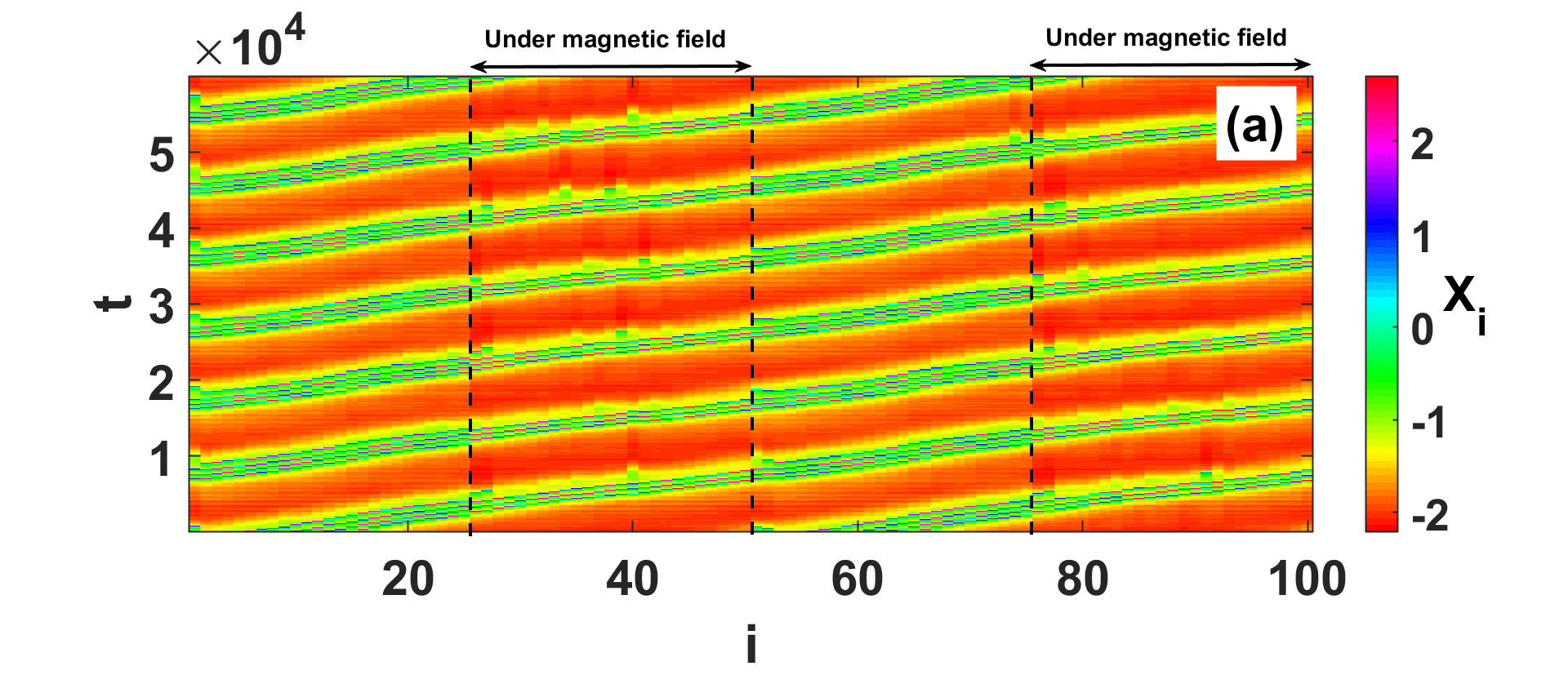}	
	\includegraphics[width=7.5cm]{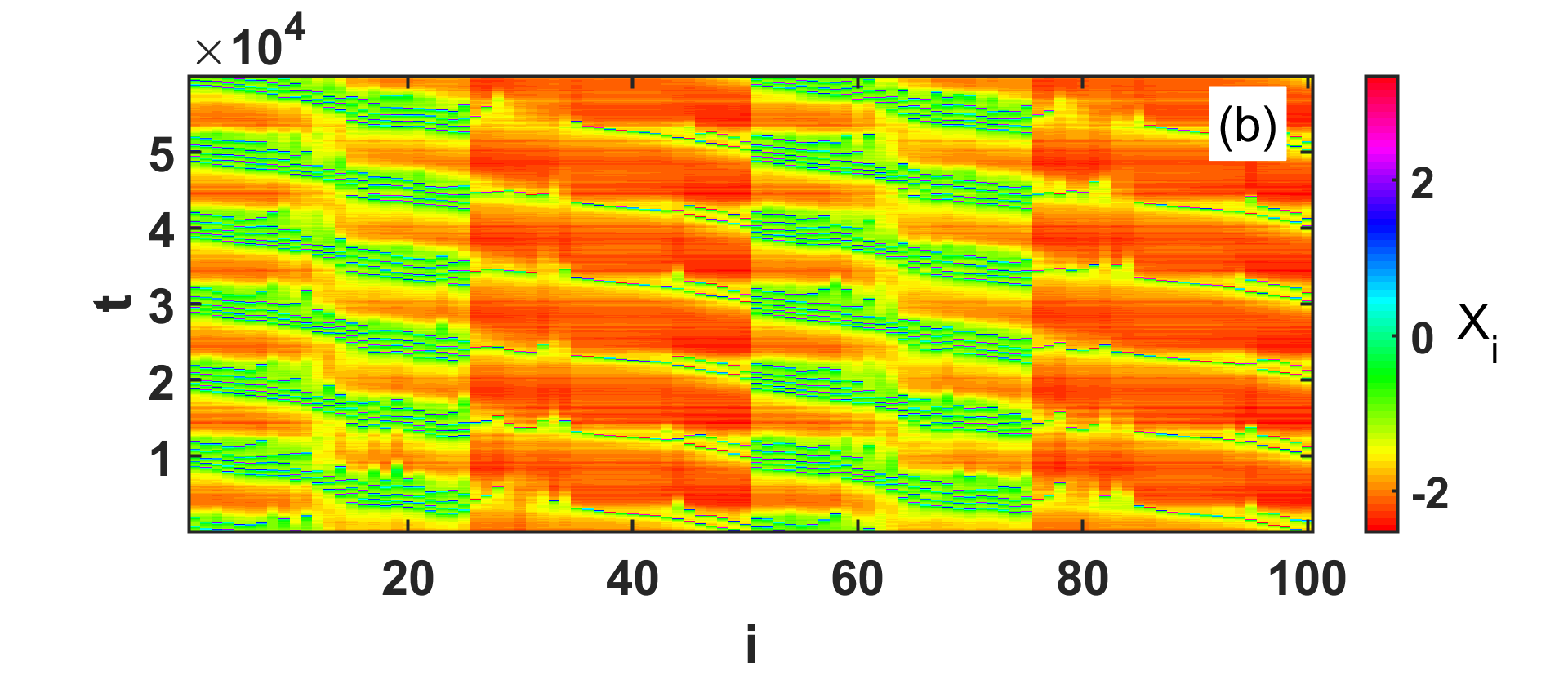}	
	\includegraphics[width=7.5cm]{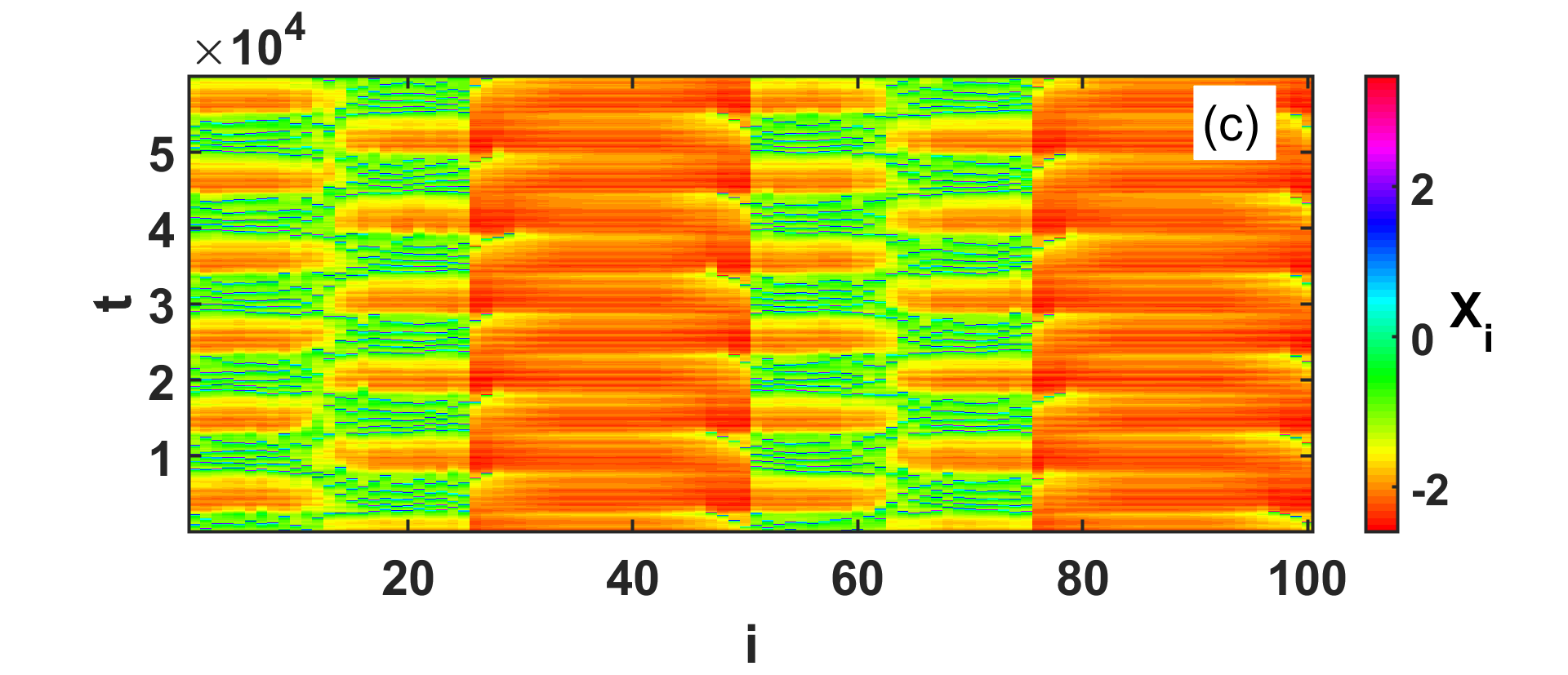}
	\caption{\label{fig.SGR7}
		Influence of a magnetic field applied on two different parts of the lattice,for traveling chimera state, for $f=0.5$: Spatiotemporal evolutions of the variables $x$ with (a) for $k_1=5$, no significant change,we see the begining of the degradation of decoherent bands; (b) multitraveling chimera state for $k_1=20$, with destruction of the decoherent bands; (c) multi-alternating  chimera state whith coherent domain for $k_1=25$. } 
\end{figure}

 We applied the same field simultaneously to two distinct regions, specifically, to the elements located between indices 25 and 50, and between indices 75 and 100. We began with the traveling chimera state. For $k_1$ values less than 5 ($k_1\leq 5$), no significant change appeared in the overall behavior of the entire network apart from a beginning of degradation of the incoherent bands in the areas subjected to the field (Fig.\ref{fig.SGR7}a). For  $k_1$ values between 5 and 20, we observed a narrowing of the incoherent bands in the areas subjected to the magnetic field until their complete disappearance, inducing a multi-traveling chimera state; thus showing a kind of depression or absence of bursts (sleep or inactive neurons). This suggests a kind of transition. (Fig.\ref{fig.SGR7}b). We also witnessed, in the area not subjected to the field, a splitting and a horizontal straightening with formation of the patterns of alternative chimera state, inducing a multi-alternating chimera state (Fig.\ref{fig.SGR7}c).  At this point, the nature of these effects is not completely clear and will be the subject of further research.

\begin{figure}[!h]	
	\includegraphics[width=7.5cm]{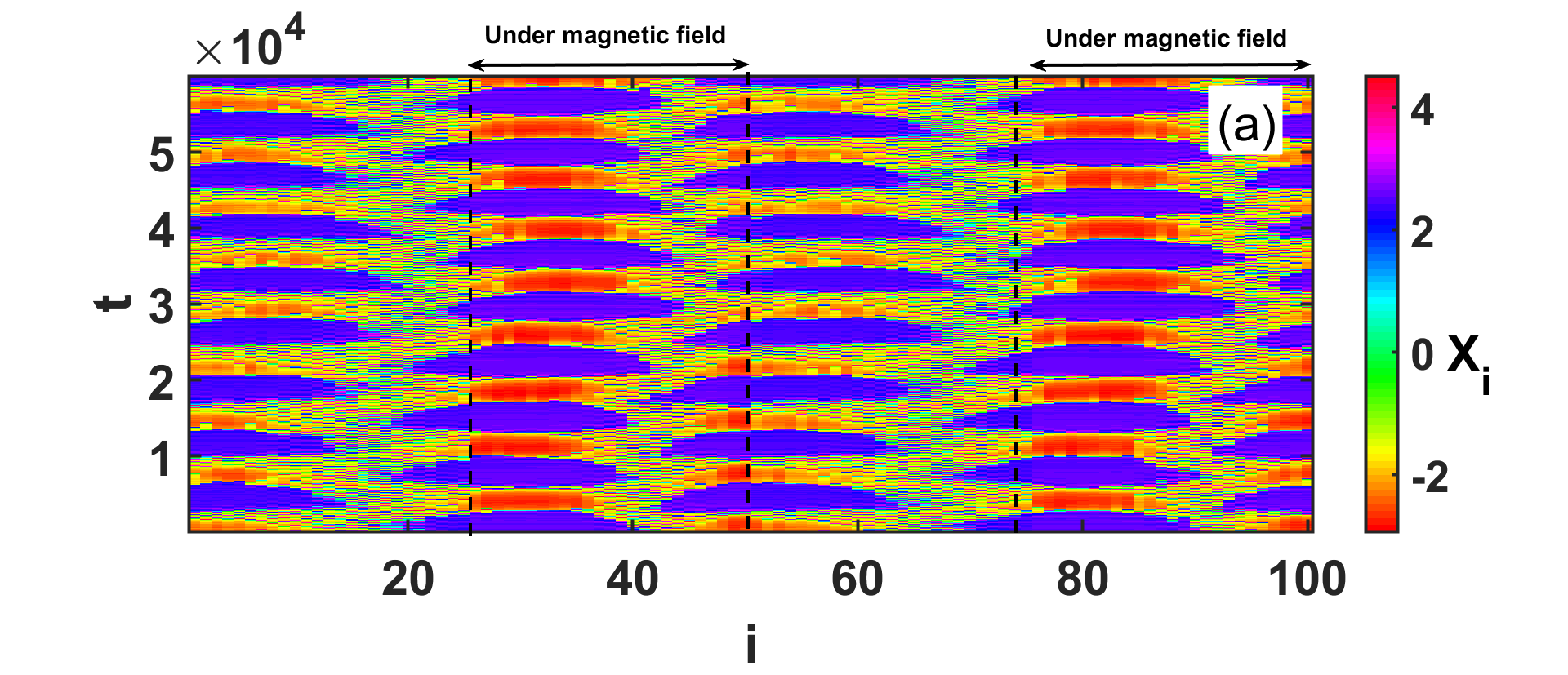}	
	\includegraphics[width=7.5cm]{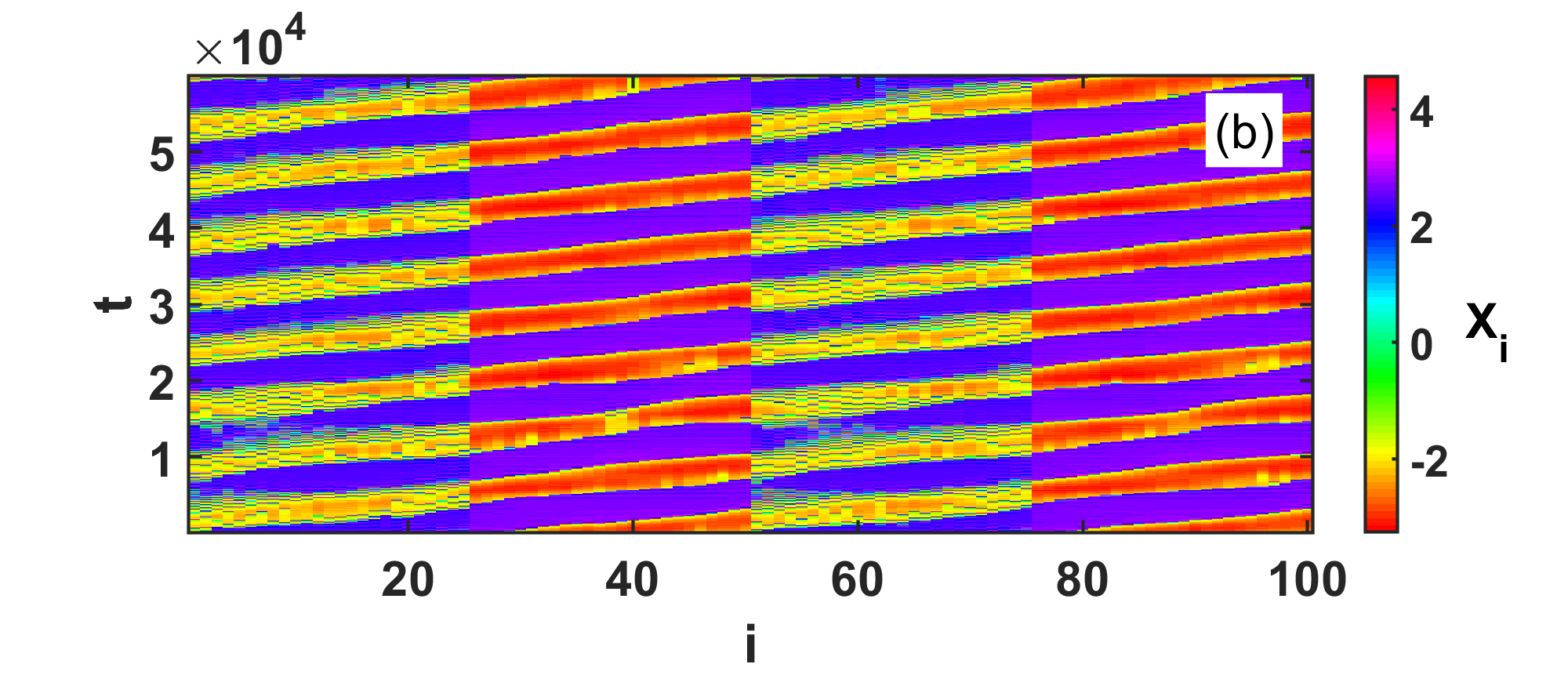}	
	\includegraphics[width=7.5cm]{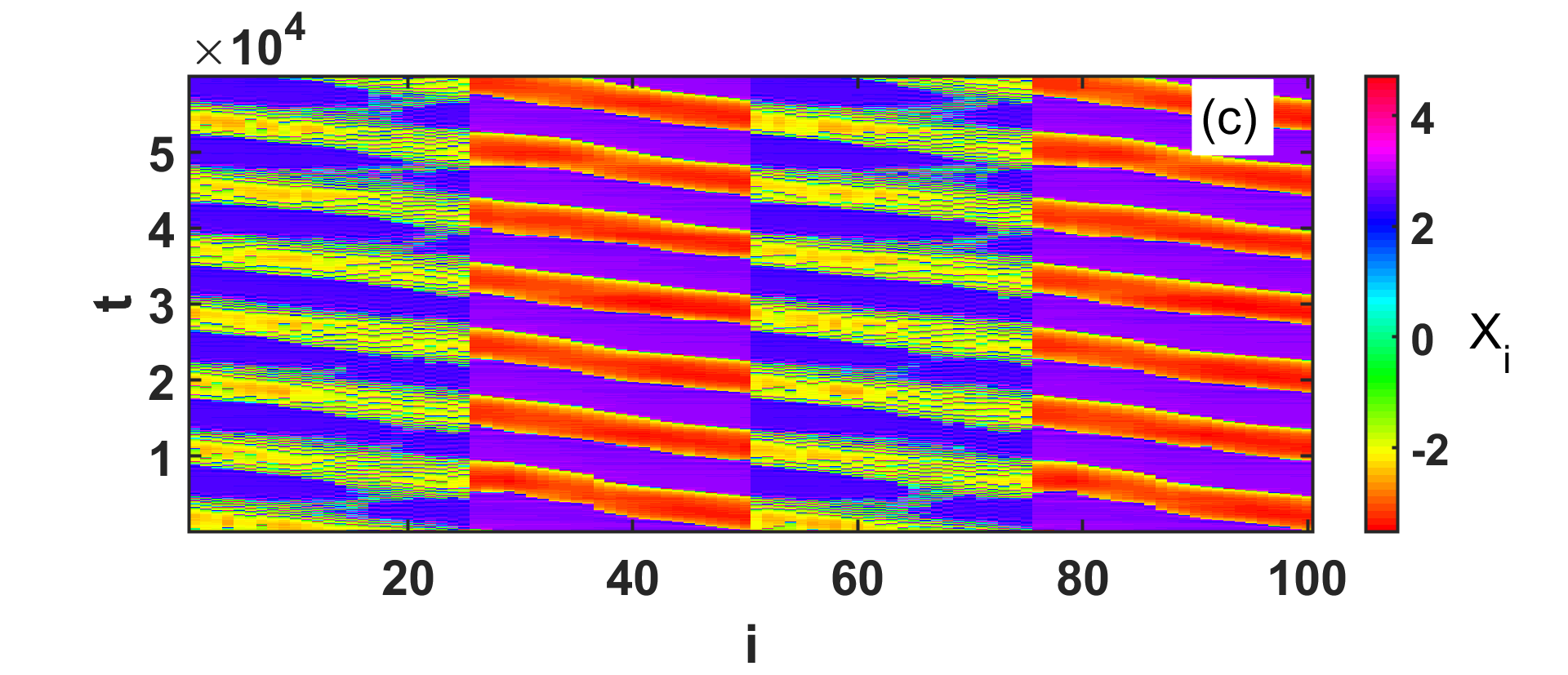}
	\caption{\label{fig.SGR8}
		Influence of a magnetic field applied on two different parts of the lattice, for traveling Multicluster Chimera state, with $f=0.5$: Spatiotemporal evolutions of the variables $x$ with (a) for $k_1=8$, no significant change, we see the the increase in the red color in these coherence zones; b) for $k_1=20$, appearance of the multitraveling chimera state or multitraveling wave; (c) change in the direction of the waves and appearance of the imperfect traveling chimera state in areas not subject to the field for $k_1=25$ } 
\end{figure}

We then apply the field to the Multicluster Chimera state. It appears that for values of $k_1$ smaller than 8, there is no major change, with the unique difference that the coherence zones under a magnetic field exhibit a qualitatively higher level of coherence than those not subjected to the field (Fig.\ref{fig.SGR8}a), which is observed by the increase in the red color in these coherence zones. Above 8 (ie $8< k_1 <  25$), the multicluster chimera state transforms into the coexistence of two phenomena: the emergence of the traveling wave in the zones subjected to the field and the traveling chimera state (Fig.\ref{fig.SGR8}b) in the zones not subjected to the field, thus giving rise to the multitraveling chimera state or multitraveling wave. As the value of $k_1$ becomes increasingly large ($k_1\geq 25$), the direction of the waves changes, and the zones exhibiting the traveling chimera state transform into the imperfect traveling chimera state (Fig.\ref{fig.SGR8}c).

\begin{figure}[!h]	
	\includegraphics[width=7.5cm]{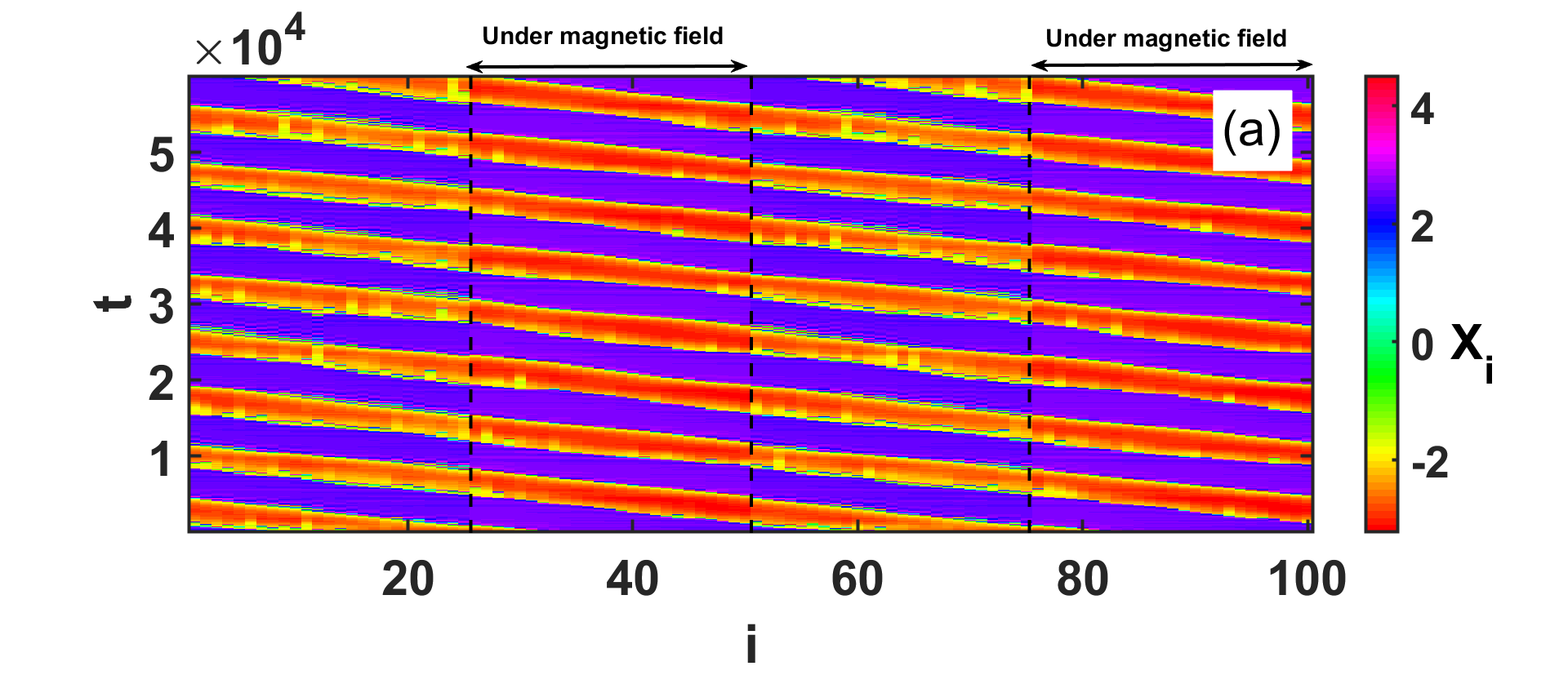}	
	\includegraphics[width=7.5cm]{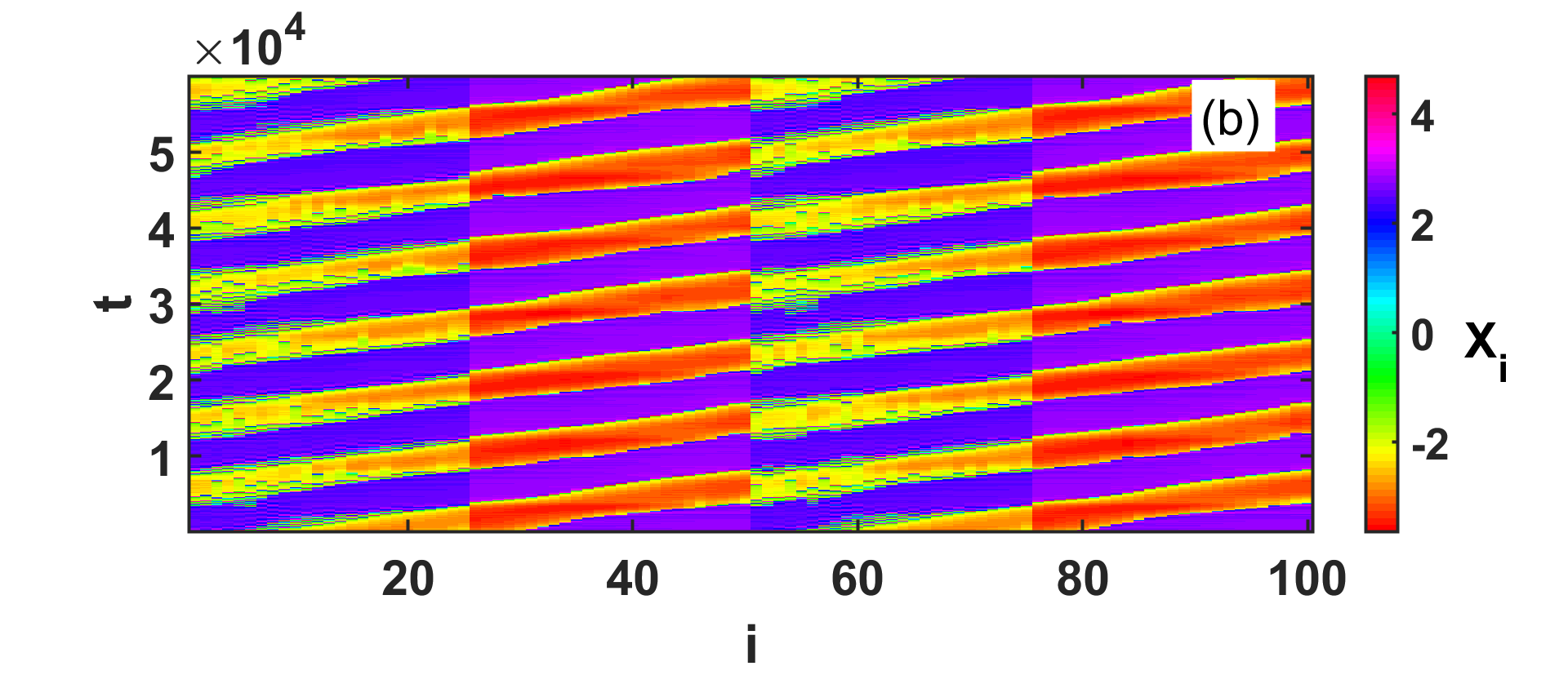}		
	\caption{\label{fig.SGR9}
		Influence of a magnetic field applied on two different parts of the lattice, for a traveling Multicluster chimera breather, with $f=0.5$: Spatiotemporal evolutions of the variable $x$ with (a) for $k_1=8$, highlighting the imperfect traveling wave; (b) for $k_1=20$, coexistence of traveling wave and the imperfect traveling chimera state. } 
\end{figure}

Finally, we apply the field to the traveling multicluster chimera breather. We observe that, for \textcolor{Brown}{$k_1$} values less than 6, the overall behavior of the network remains almost unchanged. When $k$ is between 6 and 20, the entire network becomes the site of the traveling wave (Fig.\ref{fig.SGR9}a). When $k$ takes values greater than 20, an imperfect traveling chimera state appears in the region not subjected to the field, while a traveling wave surges in the region subjected to the field (Fig.\ref{fig.SGR9}b).

\begin{figure}[!h]	
	\includegraphics[width=7.5cm]{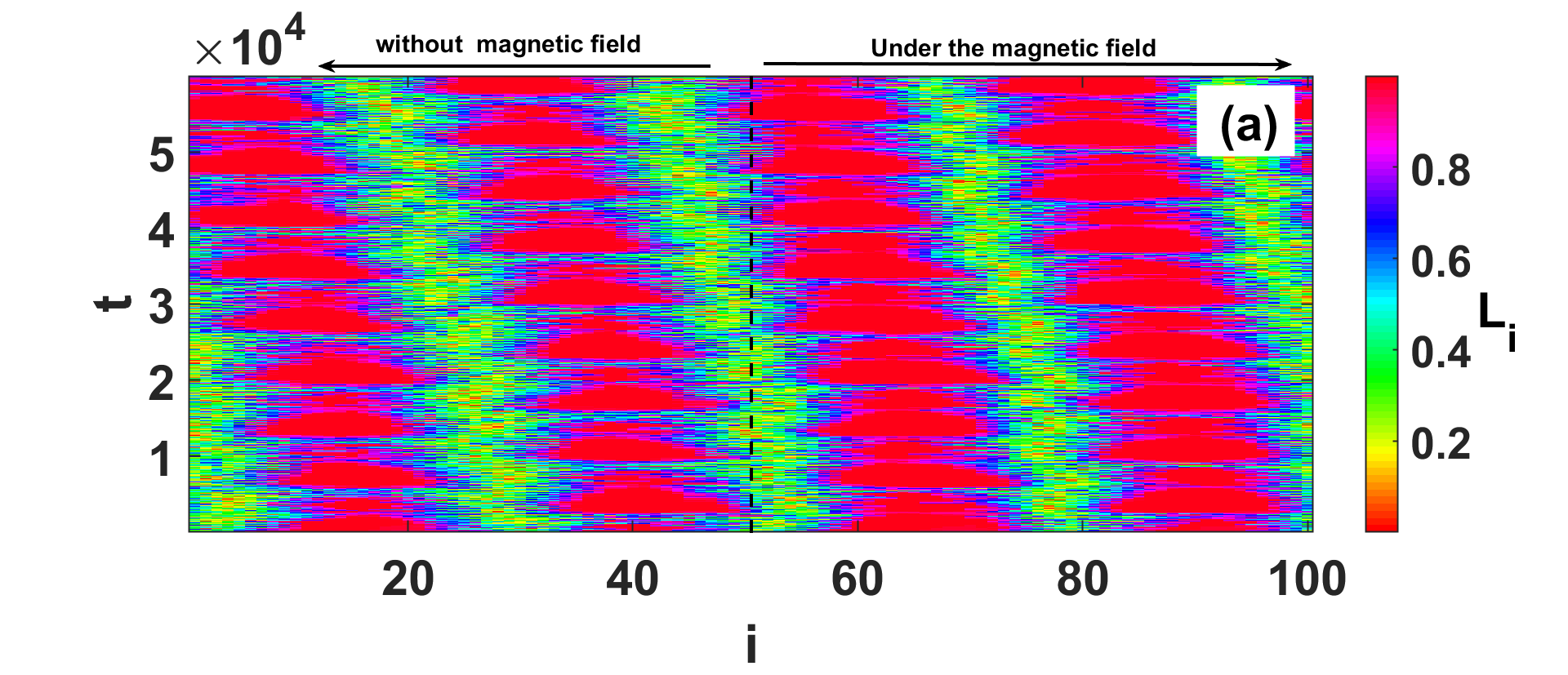}	
	\includegraphics[width=7.5cm]{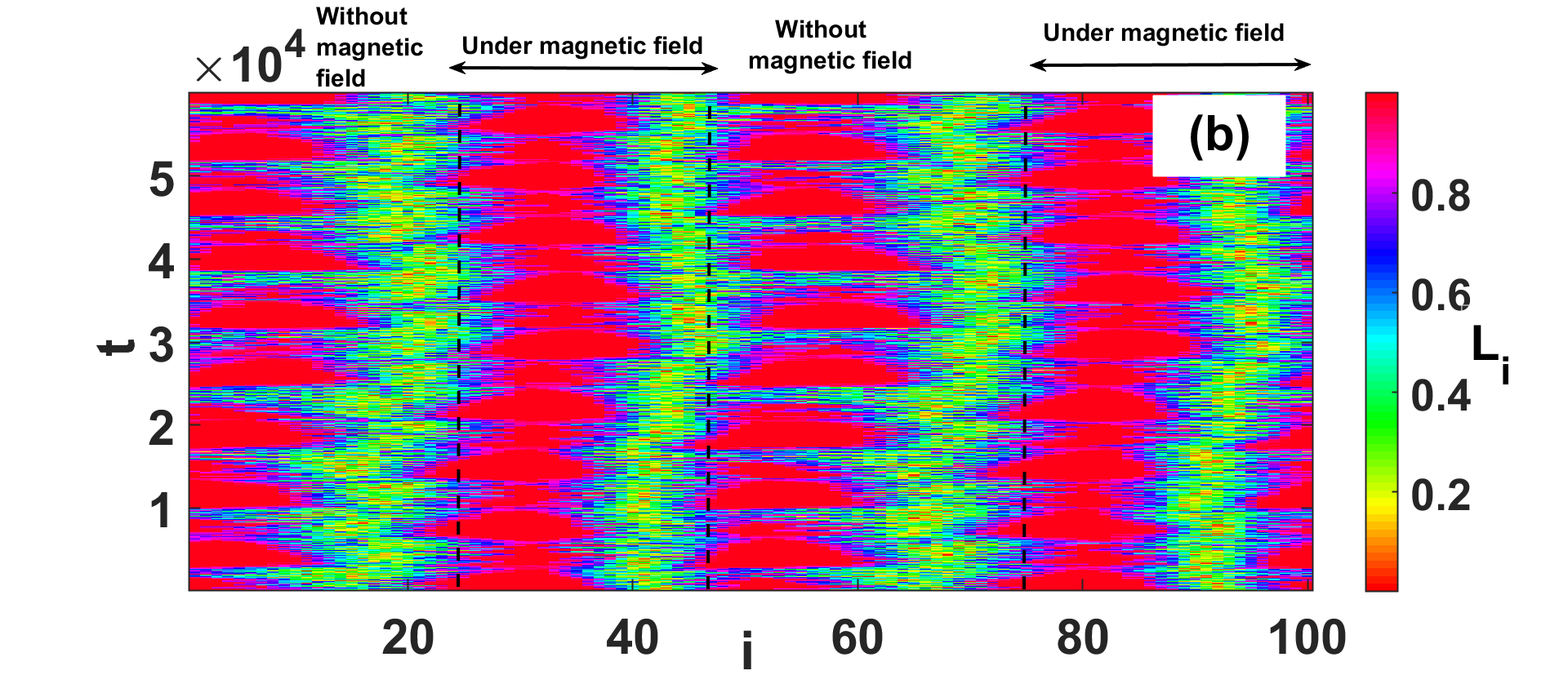}	
	\caption{\label{fig.SGR10}
		Local order parameter $L_i$ corresponding respectively to: a) the case of partial application of the magnetic field to the second half of the network (Fig. 6a), and b) the case of applying the magnetic field to two symmetric locations on the network (Fig. 8a), for the traveling multicluster chimera state. } 
\end{figure}

\begin{figure}[!h]	
	\includegraphics[width=7.5cm]{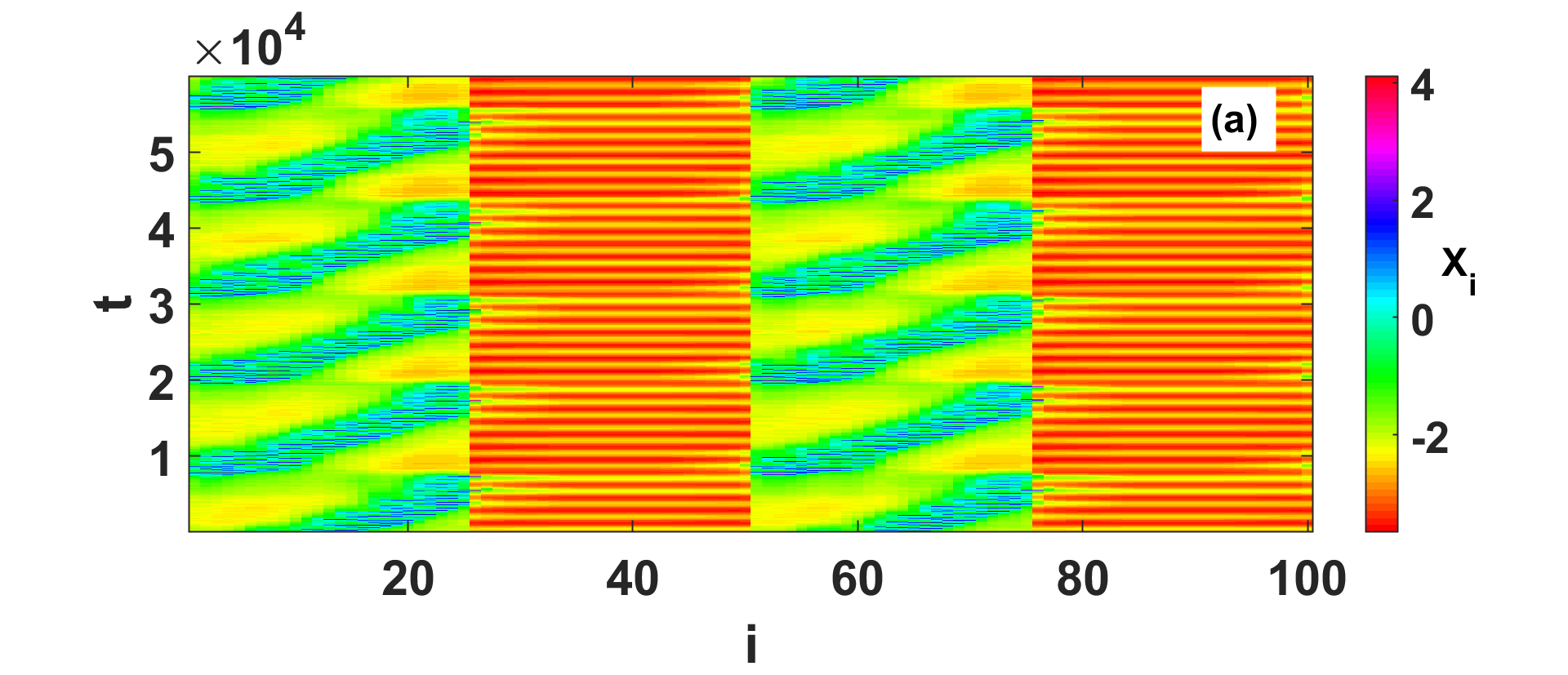}	
	\includegraphics[width=7.5cm]{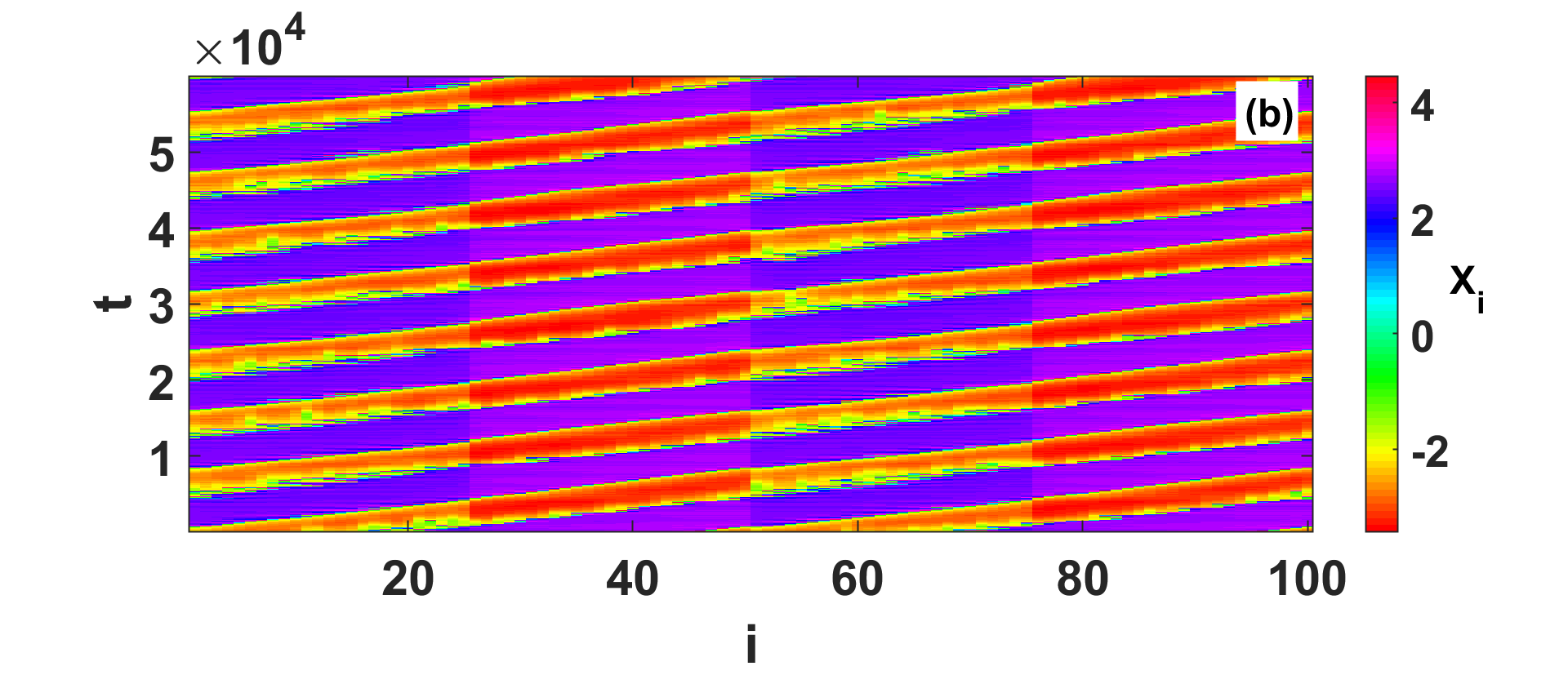}		
	\caption{\label{fig.SGR11}
		Sensitivity to variations in the electrical and chemical coupling coefficients: Sensitivity to variations in the electrical and chemical coupling coefficients: a) transition from the multi-alternating chimera state to the multi-traveling chimera state for $k_3=1$, $k_4=11$, and $I=3.5$; b) transition from the multi-traveling chimera state to the imperfect traveling wave for $k_3=1$, $k_4=10$, and $I=3.5$. } 
\end{figure}

To substantiate the observations made in Fig.\ref{fig.SGR6}a and Fig.\ref{fig.SGR8}a regarding applying the magnetic field to half the lattice and to two symmetrical locations, respectively, we plot the local order parameter $L_i$ (\cite{Simo}) for these two situations. The results (Fig.\ref{fig.SGR10}) show a continuous intensification of the red color in the regions subjected to the field, indicating that the order parameter tends toward 1. This confirms the color enhancement, thereby reflecting an increase in the degree of coherence within the areas exposed to the field.\\
We consider the multialternating chimera state obtained in Fig.\ref{fig.SGR7}c and set the chemical and electrical coupling parameters to 1 and 11, respectively (ie $k_3=1$, $k_4=11$). This results in the transformation of the multi-alternating chimera state into a multi-traveling chimera state, as shown in Fig.\ref{fig.SGR11}a. When we perform the same procedure (ie $k_3=1$, $k_4=10$) for the multi-traveling chimera state presented in Fig.\ref{fig.SGR8}c, we observe that it disappears, giving way to an imperfect traveling wave, as shown in Fig.\ref{fig.SGR11}b for the same parameter values. This demonstrates the sensitivity of both the multi-alternating chimera state and the multi-traveling chimera state to variations in the chemical and electrical coupling parameters.

\begin{figure}[!h]	
	\includegraphics[width=7.5cm]{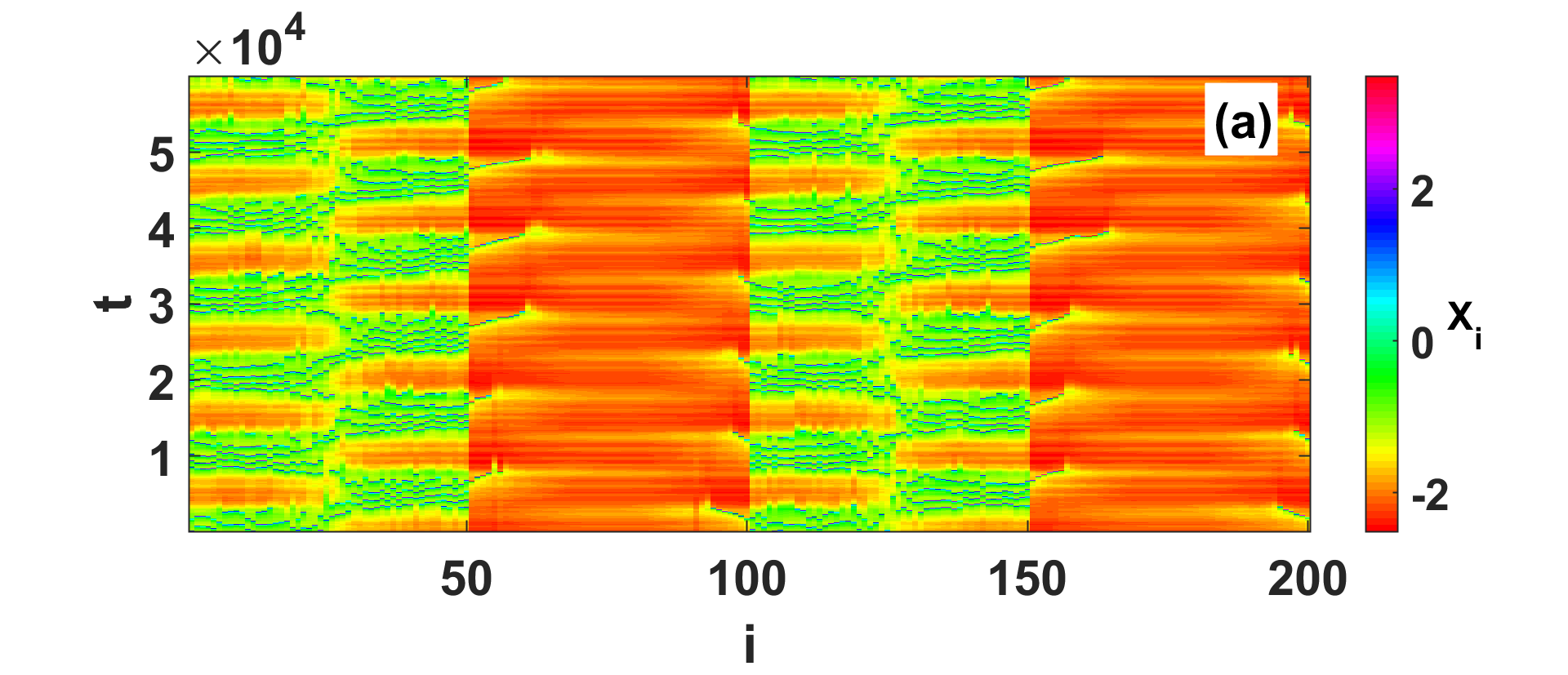}	
	\includegraphics[width=7.5cm]{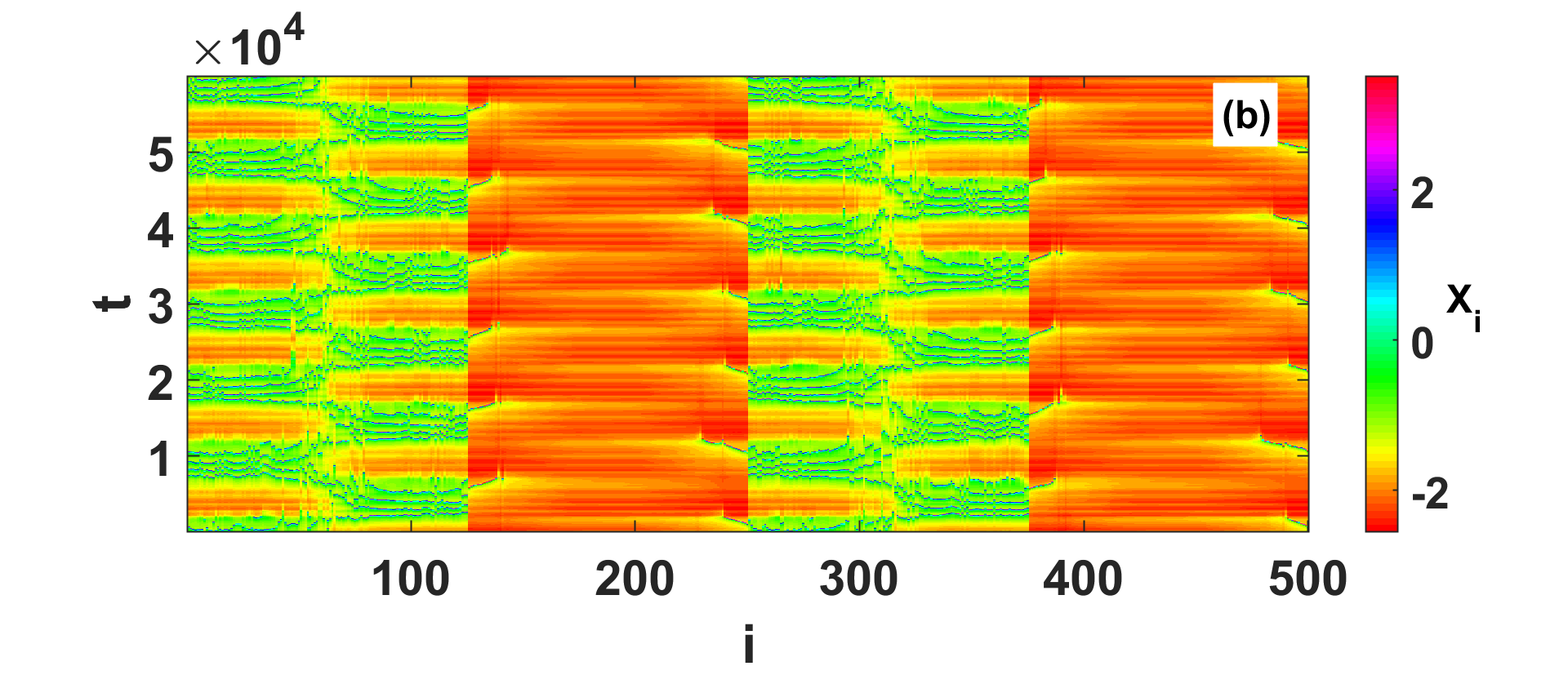}
    \includegraphics[width=7.5cm]{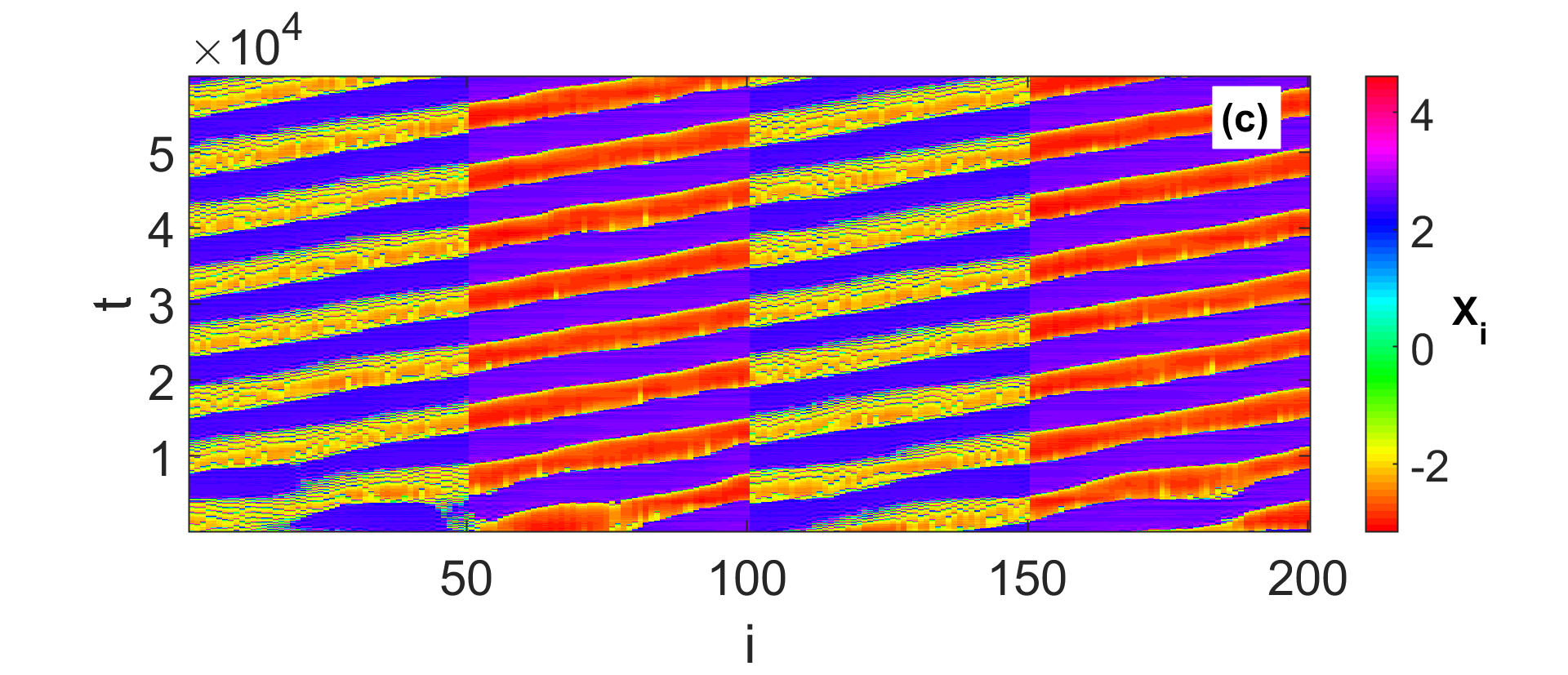}	
	\includegraphics[width=7.5cm]{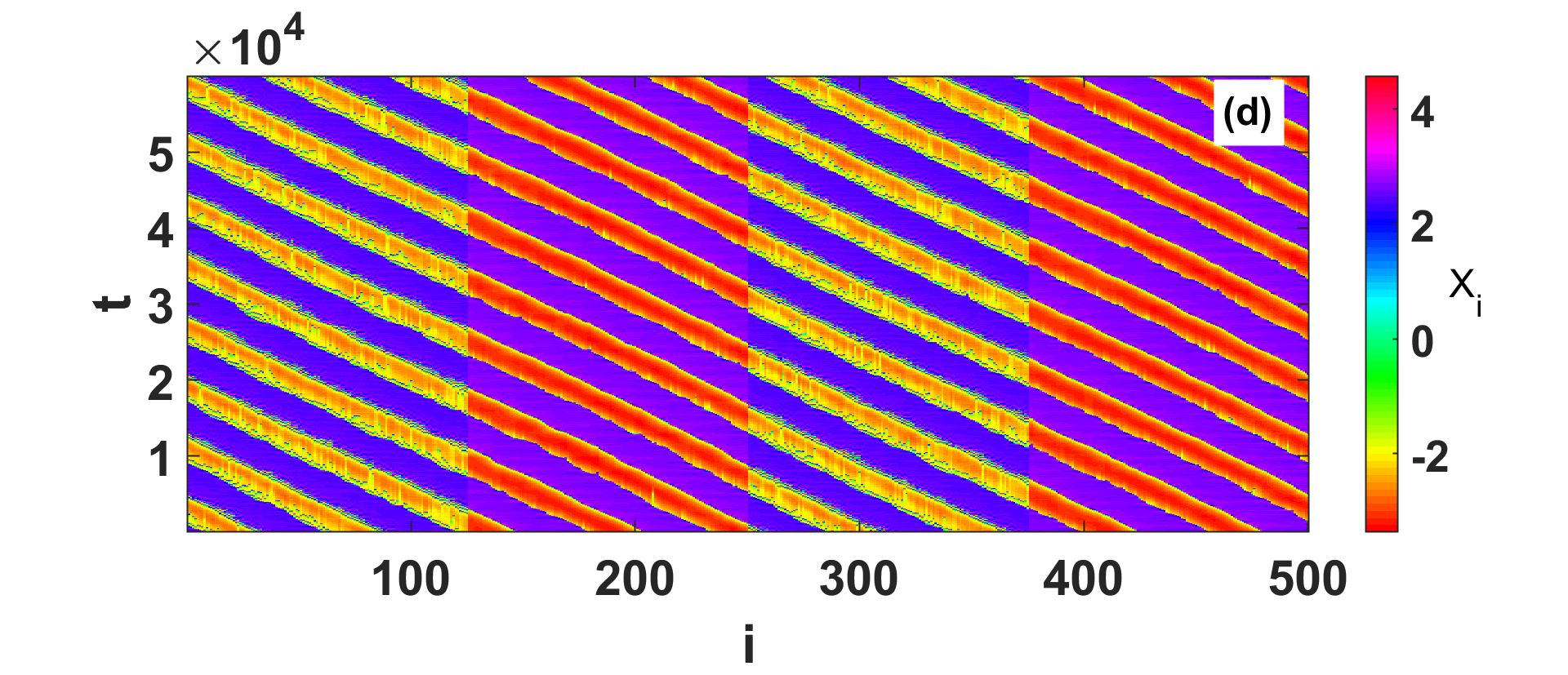}
	\caption{\label{fig.SGR12}
		Variation of the size M of the network: a) multialternating chimera state for M=200; b) multialternating chimera state for M=500; c) multitraveling chimera state for M= 200; and d) multitraveling chimera state for M=500. } 
\end{figure}

To verify the generality of the phenomena observed with respect to the number of network elements $M$, we repeat the processes shown in Fig.\ref{fig.SGR7}c and Fig.\ref{fig.SGR8}b for two different, randomly chosen values of $M$. We apply the field using the same proportions as before: by dividing the network into four parts, each containing an equal number of elements. Upon application of the magnetic field under the same conditions, the formation of a multialternating chimera state is observed, as shown in Figs.\ref{fig.SGR12}a and b, and the formation of a multitraveling chimera state, as shown in Figs.\ref{fig.SGR12}c and d.

\subsection{Frequency analysis}
To provide a physical interpretation of the observed phenomena, we proceed in two ways: first, for each of the preceding cases, we vary the frequency of the external magnetic field to determine its influence; subsequently, we perform a frequency analysis of the waveforms of the network elements.\\
When we vary the external field frequency $f_{mth}$ for the aforementioned cases, it appears that, for each scenario, a threshold frequency $f_{mth}$ exists, at which the corresponding chimera state emerges. These frequencies are listed in Tab.\ref{tabI} below.\\

\begin{table}[htbp]
\centering
\caption{Threshold frequencies} 
\label{tabI} 
\vspace{0.5 em} 

\begin{tabular}{|c|c|c|c|c|c|c|c|}
\hline
Fig. & Fig.\ref{fig.SGR2}c & Fig.\ref{fig.SGR4}a & Fig.\ref{fig.SGR5}c & Fig.\ref{fig.SGR6}b & Fig.\ref{fig.SGR7}c & Fig.\ref{fig.SGR8}c & Fig.\ref{fig.SGR9}b   \\ \hline
$f_{mth}$ & 0.07 & 0.1 & 0.2 & 0.04 & 0.03 & 0.1 & 0.05 \\ \hline
\end{tabular}
\end{table}

\begin{figure}[!h]	
	\includegraphics[width=7.5cm]{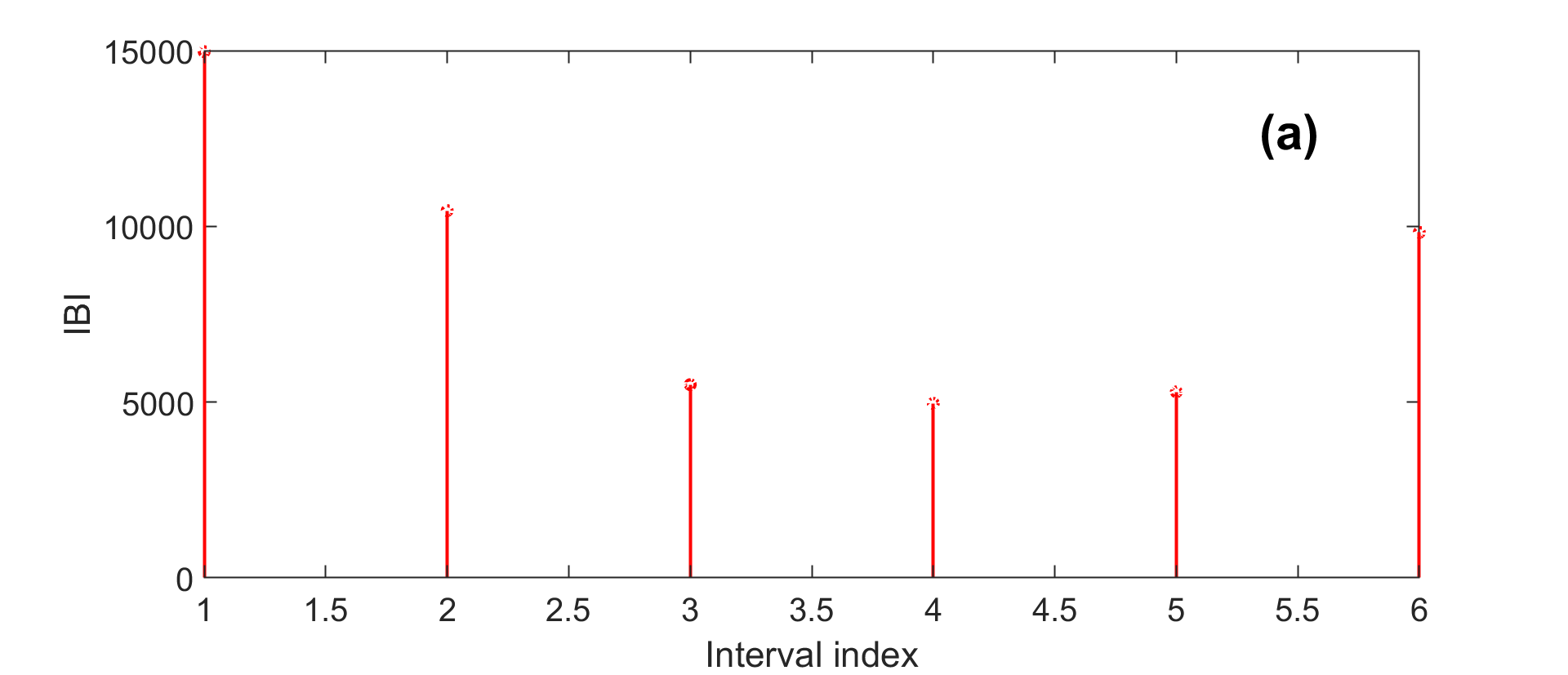}	
	\includegraphics[width=7.5cm]{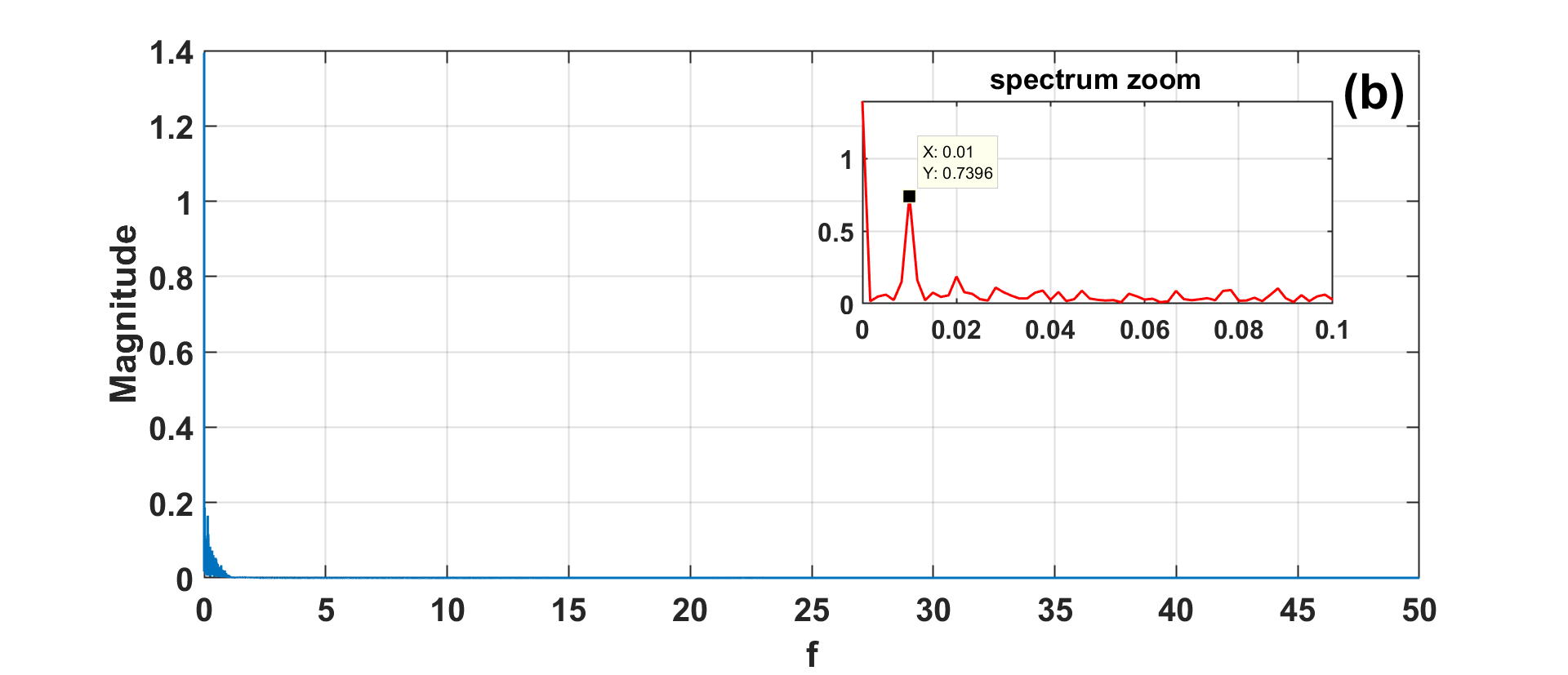}		
	\caption{\label{fig.SGR13}
		Frequency analysis for the case of the imperfect traveling chimera state shown in Fig.\ref{fig.SGR2}(c), illustrated by (a) the interburst interval, which yields a burst frequency of $f_{bi} = 0.0139$, and (b) a Fourier spectrum showing that the fundamental frequency $f_0$ is on the order of $0.01$, as shown in the zoomed-in view.  This illustration corresponds to neuron 85 and is representative of the other neurons in the network. } 
\end{figure}

We are particularly interested in the average burst frequency, $f_{bi}$, in comparison with the frequency of the applied magnetic field, $f_m$. To evaluate $f_{bi}$, we use the following analytical expression, including the inter-burst interval (IBI) \cite{FrequencyLA, WangH}:
\begin{equation}
f_bi = \frac{1}{K_i}
\sum_{k=1}^{K_i}
\frac{1}{\tau_i^{k+1} - \tau_i^k}
\end{equation}
where:
\begin{itemize}
    \item $f_bi$ is the average burst frequency of neuron $i$;
    \item $K_i$, the number of inter-burst intervals considered for neuron $i$;
    \item $\tau_i^k$, the onset time of the $k$-th burst of neuron $i$;
    \item $\tau_i^{k+1}$, the onset time of the $(k+1)$-th burst of neuron $i$;
    \item $\tau_i^{k+1} - \tau_i^k$, the inter-burst interval (IBI) between two consecutive bursts.
\end{itemize}
We also performed a Fourier spectrum analysis to extract the fundamental frequency $f_0i$ of each waveform under study. An illustrative example concerning the case of Fig.\ref{fig.SGR2}(c) is shown in Fig.\ref{fig.SGR13}.\\
In Fig.\ref{fig.SGR13}(a), we present the interburst interval (IBI) for neuron 85 of the network, which is subjected to a magnetic field that gives rise to an imperfect traveling chimera state.It shows exactly 6 intervals of varying widths, corresponding to the number of bursts observed in the waveforms of any network element under these conditions. This result is obtained using a trigger threshold of $-1.7$, chosen specifically to satisfy $\dot{x} > 0$, resulting in a burst frequency of $f_{bi} = 0.0139$. Furthermore, when examining the wave frequency, it appears to be very close to the burst frequency, remaining fixed at $f_{oi} = 0.01$, as shown in the zoomed-in portion of Fig.\ref{fig.SGR13}(b), which displays the Fourier spectrum.
\\

\begin{table}[htbp]
\centering
\caption{Burst frequencies $f_{bi}$ and fundamental frequencies $f_{oi}$} 
\label{tabII} 
\vspace{0.5 em}

\begin{tabular}{|c|c|c|c|c|c|c|c|}
\hline
Fig. & Fig.\ref{fig.SGR2}c & Fig.\ref{fig.SGR4}a & Fig.\ref{fig.SGR5}c & Fig.\ref{fig.SGR6}b & Fig.\ref{fig.SGR7}c & Fig.\ref{fig.SGR8}c & Fig.\ref{fig.SGR9}b   \\ \hline
$f_{bi}$ & 0.0139 & 0.0144 & 0.0141 & 0.0134 & -- & 0.0121 & 0.0115 \\ \hline
$f_{oi}$ & 0.01 & 0.02 & 0.01 & 0.01 & 0.02 & 0.01 & 0.01  \\\hline
\end{tabular}
\end{table}
The same procedure is carried out for the remaining scenarios(see Tab.\ref{tabII}). It is worth noting that, despite the fundamental frequencies $f_{oi}$ being very close to the burst frequencies $f_{bi}$, there is a frequency locking between the frequency of the external field ($f_m = 0.5$) and the fundamental frequencies of the neurons subjected to that field. It should be noted that the absence of a burst frequency in the case shown in Fig.\ref{fig.SGR7}c is due to the lack of bursts in the region exposed to the field; we are dealing solely with oscillations.

\section{Conclusion}

We studied the influence of a magnetic field on a one-dimensional HR-type neuronal network. Lv and Ma initially proposed the model \cite{Lv}, which accounts for magnetic coupling. This work follows on from the work carried out by Simo et al. \cite{Simo}, in which they highlighted the influence of the electric field. In the present case, we consider three main phenomena that have attracted our attention in previous studies, namely: the traveling chimera state, the traveling multicluster chimera state, and the traveling multicluster chimera breather. Since all three phenomena were obtained only under chemical coupling, we placed our network under the same conditions to study the influence of an external magnetic field.
Our results show that the magnetic field can reorganize coherent and incoherent neuronal populations, induce transitions between chimera and traveling wave states, and generate multi-traveling and multi-alternating chimera states when applied to selected network regions.

We compare the present results with our previous investigation of electric-field effects on the same class of Hindmarsh–Rose neuronal networks [44]. In both studies, an external field modifies the collective dynamics and spatial organization of neuronal activity. However, the magnetic-field-induced transitions observed here exhibit specific features, including the enhancement of coherence in the field-exposed regions and the emergence of multitraveling and multialternating chimera states under spatially localized field application.
\\
A further distinction concerns oscillation death. While oscillation death was reported in the electrically driven network, we observed no such behavior in the regions exposed to the magnetic field within the parameter ranges investigated in the present study. Instead, these regions can retain oscillatory activity while evolving toward more coherent patterns or traveling waves. This contrast points to potentially different dynamical responses to electric and magnetic forcing. Nevertheless, a systematic quantitative comparison requires examining the two field models under matched parameter conditions using common observables, such as the local order parameter, the fraction of incoherent neurons, neuronal burst and fundamental frequencies, and the occurrence of oscillation death. Extending the analysis to the simultaneous application of electric and magnetic fields is a natural direction for future work.

\section*{Acknowledgements}
P.L. and H.A.C. are thankful for ICTP-SAIFR and {Fundaç\~ao} de Amparo à Pesquisa do Estado de {S\~ao} Paulo (FAPESP): 2021/14335-0 for partial support.

\end{document}